\documentclass[smallextended]{svjour3}
\usepackage{url}
\usepackage{graphicx}
\usepackage{amssymb}
\usepackage{natbib}%[numbers,sort&compress]
\usepackage{aas_macros}
\usepackage{lscape} 
\usepackage{hhline}
\usepackage{xcolor}
\usepackage{tcolorbox}
\usepackage{colortbl}

\newcommand{\RowColor}{\rowcolor{red!50} \cellcolor{white}}

\begin{document}

\title{\emph{Ariel Planetary Interiors} White Paper}
%
%\author[1]{Ravit Helled}
\author{Ravit Helled, Stephanie Werner, Caroline Dorn, Tristan Guillot,  Masahiro Ikoma,  Yuichi Ito, Mihkel Kama, Tim Lichtenberg, Yamila Miguel, Oliver Shorttle, Paul J. Tackley, Diana Valencia, Allona Vazan} 
\authorrunning{Helled et al.}
%\author[2] {Stephanie C. Werner}
%\author[3] {Caroline Dorn, Tristan Guillot,  Masahiro Ikoma,  Yuichi Ito, Tim Lichtenberg,Yamila Miguel, Diana Valencia, Allona Vazan}
%\affil[1]{Institute of Space Astrophysics and Planetology INAF-IAPS}
\institute{ 
            Helled, R., Dorn, C., \at
            Institute for Computational Science, University of Zurich, Winterthurerstr. 190, CH-8057 Zurich, Switzerland \email{rhelled@physik.uzh.ch}
            \and
            Werner, S. \at
            Department of Geosciences, 
            University of Oslo, PO 1028 Blindern, 0315 Oslo, Norway
            \and
            Guillot, T. \at 
            Observatoire de la Côte d’Azur / CNRS, 6304 Nice Cedex 4, France
            \and 
            Ikoma, M. \at 
            The University of Tokyo, Department of Earth and Planetary Science, Tokyo 113-0033, Japan 
            \and
            Ito, Y., \at 
            Department of Physics and Astronomy, University College London, London, WC1E 6BT, United Kingdom
            \and
            Kama, M. \at
            Tartu Observatory, University of Tartu, Observatooriumi 1, 61602, T\~{o}ravere, Estonia
            \and 
            Lichtenberg, T. \at 
            Atmospheric, Oceanic and Planetary Physics,  Department of Physics, University of Oxford, United Kingdom
            \and
           Miguel, Y.\at
           Leiden Observatory, University of Leiden, Niels Bohrweg 2, 2333CA Leiden, The  Netherlands,\\
          SRON Netherlands Institute for Space Research , Sorbonnelaan 2, NL-3584 CA Utrecht, the Netherlands
            \and
            Shorttle, O. \at
            Institute of Astronomy, University of Cambridge, Cambridge, UK
            \and
            Tackley, P. J. \at
            Institute of Geophysics, ETH Zurich, 8092 Zurich, Switzerland
            \and 
            Valencia, D. \at 
            University of Toronto, Canada
            \and 
            Vazan, A. \at 
            Department of Natural Sciences, the Open University of Israel, Israel
}
%\affil[1]{University of Zurich}
%\affil[2]{Oslo Univ.}
%\affil[3]{Please put your affiliations...}
%

\maketitle
\begin{abstract}
The recently adopted {\it Ariel} ESA mission will measure the atmospheric composition of a large number of exoplanets. This information will then be used to better constrain planetary bulk compositions. While the connection between the composition of a planetary atmosphere and the bulk interior is still being investigated, the combination of the atmospheric composition with the measured mass and radius of exoplanets will push the field of exoplanet characterisation to  the next level, and provide new insights of the nature of planets in our galaxy.  
In this white paper, we outline the ongoing activities of the interior working group of the {\it Ariel} mission, and list the desirable theoretical developments as well as the challenges in linking planetary atmospheres, bulk composition and interior structure.  
%Please provide an abstract of no more than 300 words. Your abstract should explain the main contributions of your article, and should not contain any material that is not included in the main text. 
\end{abstract}


% Compatibility w/ ADS journal abbreviations (TL)
%\include{journal_abbr.txt}

%
%\begin{document}

%\flushbottom
\maketitle
%\thispagestyle{empty}

% Comments from discussion: 
% Jon's contribution?  
% Discuss the expected difference of the atmospheric and deep interior composition. 
% Targets: very dense planets, inflated JUpiters - full light curve ohomic dissipation: Thongern and Fortney 
% condensation of iron... 
% elemnts: refractory material abundances - metallicity 
% very dense giant planets - what is their atmospheric composition? are the heavies mostly in the deep interior? 
% what we expect from formation and evolution?
% what do we know from the solar system?
%Ariel has been selected as the next ESA medium-class science mission and is due for launch in 2028. During its 4 yr mission, Ariel aims to observe ∼1000 exoplanets ranging from Jupiters and Neptunes down to super-Earth size in the visible and the infrared with its meter-class telescope. 
\clearpage
\section{Introduction}
The Atmospheric Remote‐sensing Infrared Exoplanet Large‐survey ({\it Ariel}) mission will measure the atmospheric composition of a large number of exoplanets with different masses and radii \citep[e.g.,][]{Tinetti+2018,2019AJ....157..242E}. %(Tinetti et al., 2017). 
A key question to be addressed by the {\it Ariel} mission is `{\it What are planets made of?}'

Understanding the connection between the bulk composition and atmospheric composition of planets and how they are linked to the planetary origin is a key topic in planetary and exoplanetary science. %Understanding the connection between the planetary origin, the bulk, and atmospheric composition is key. 
Determining the atmospheric composition of planets is critical for constraining the planetary bulk composition, which can then be linked to its formation process and  planetary evolution \citep[e.g.,][]{2018ExA....46...45T}. Determining the atmospheric composition of many exoplanets will provide an additional constraint to structure models, and can be used to put limits on elemental abundances in the deep interiors of planets. 
While it is not possible to uniquely determine planetary composition and internal structure from remote measurements, the atmospheric composition adds another piece of information and can break some of the degeneracy.    
%which could still be different from that of the bulk, this already reduces the uncertainty linked to the choice of the type of heavy elements to consider in modeling the interior (by fixing the C/O ratio for example).
In this regard, a clear strength of the {\it Ariel} mission is that it will provide statistics. A large enough sample of planetary atmospheres can be used to better understand the trends, in particular, how the planetary composition depends on the planetary mass, the orbital period, and the stellar properties such as stellar type, metallicity, and age. 
%lthough exoplanetary characterization of individual objects is 

%\begin{figure}[h!]
%\floatbox[{\capbeside\thisfloatsetup{capbesideposition={left,top},capbesidewidth=4cm}}]{figure}[\FBwidth]
%\vskip -8pt
%\vspace{-0.8cm}
%\centering
%{\includegraphics[scale=.56]{slices_JS.pdf}}
%{\includegraphics[scale=.0454]{fig1_new.jpg}}
%{ \caption{\small A sketch demonstrating the connection between planet formation, evolution, and internal structure. The formation process determines the planetary composition and primordial internal structure as well as the thermal state. This, determines the long-term evolution of the planet which could include a change of the internal structure due to the redistribution of elements and the subsequent evolution. As a result, the current-state interior must be interpreted in terms of these three aspects and their interplay. }}
%\end{figure}

The three aspects of formation, evolution and internal structure are inter-connected. The formation environment and epoch determine the total availability of each chemical element.  The formation process determines the primordial internal structure and thermal state of the planet.  This determines the heat transport mechanism as well as the potential re-distribution of the materials, and the planetary long-term evolution (contraction and cooling rate). The planetary evolution then determines current-state internal structure. In particular, the interaction between the atmosphere and the deep interior. The long-term evolution could be responsible for the formation of secondary atmospheres as well as for atmospheric loss. Therefore, in order to link the planetary internal structure and to understand the connection between the atmosphere and the interior today, a good understanding of the planetary origin and evolution is required.

From a planetary composition perspective the questions that will be addressed with {\it Ariel} include:
%As a result, several questions regarding the planetary interior and bulk composition will be addressed. These include the following:  \\
\begin{itemize}
 \item[$\bullet$] What can the M-R relation together with atmospheric measurements tell us about the planet's bulk composition? 
 \item[$\bullet$] Under what conditions does the atmospheric composition represent the composition of the deep interior?
%- What elements in the atmospheres can be linked to geophysical activity and habitability? \\
 \item[$\bullet$]  Is atmospheric composition able to distinguish planetary archetypes, such as mini-Neptunes versus super-Earths? 
 \item[$\bullet$] How can we use the knowledge of exoplanets to better understand our own planetary system and vice versa?
 \item[$\bullet$] How do we extract the similarities/differences between solar system (terrestrial) planets and exoplanets from atmospheric element abundances? 
 \item[$\bullet$]  How do planetary atmospheres in hot conditions evolve?
% \item[\bullet]  What are the potential diagnostics for Ariel? Is there a specific property/element that is particularly useful? 
 %\item[\bullet] Are there specific targets in terms of mass, distance to the star etc that Ariel should concentrate on? 
\end{itemize}

Not only is connecting planetary atmospheric compositions with bulk compositions challenging, but there will also be clear differences between various planetary types (mass, orbital period, etc.), and as discussed above, with a planet's formation history, and subsequent evolution and internal structure. Key questions to be answered by {\it Ariel} with the suggested targets are summarised in Table 1. 
The colours indicate whether the question addresses composition (red), evolution (blue) or origin (green), or a combination of these aspects. 
The first three questions are fundamental: {\it ``What are exoplanets made of"? ``How do exoplanets form?" and ``How do exoplanets evolve"?}
These questions are expected to remain open for a few decades, and even with {\it Ariel} and other future missions, unique and clear answers might not be available. However, more specific questions, as listed in Table~\ref{table}, can slowly be answered providing a more complete understanding of planets and the connection between composition, formation and evolution. Since the three aspects are linked, often answering a question reflects on the three of them. As a result, many of the questions listed in the table involve more than one colour representing the three fundamental questions. 
%We also indicate the targets to be observed for each of the questions (third column). 

%when one aims to connect the atmospheric composition to the bulk. Terrestrial planets 
\begin{landscape}

\begin{table}
%\centering 
\begin{flushleft}
\footnotesize{
\begin{tabular}{| l  |l | l | l | l | l  |}
\hline
                  &  &  & {\bf Question} & {\bf Ariel Measurements}  & {\bf Targets}  
\\ \hline
\RowColor       \cellcolor{red} & \cellcolor{white} & \cellcolor{white} & \textcolor{red}{What are exoplanets made of?} \cellcolor{white} &  \tiny{Combination of atmospheric spectra (atmospheric composition), lightcurves} \cellcolor{white} & \cellcolor{white} \tiny{super-Earths to giant planets} 
\\ 
\RowColor       \cellcolor{red} & \cellcolor{white} & \cellcolor{white} &  \cellcolor{white} &  \tiny{(atmospheric boundary conditions and evolution), and theoretical models} \cellcolor{white} & \cellcolor{white}  \\
\hline

\RowColor       \cellcolor{white} & \cellcolor{white} & \cellcolor{green} & \textcolor{green}{How do exoplanets form?} \cellcolor{white} &  \tiny{Lightcurves (atmospheric boundary conditions, day-night temperature variations),} \cellcolor{white} & \cellcolor{white} \tiny{Intermediate-mass and gas giant}
\\ 
\RowColor       \cellcolor{white} & \cellcolor{white} & \cellcolor{green} &  \cellcolor{white} &  \tiny{spectra (clouds, composition, presence of condensates) and theoretical models} \cellcolor{white} & \cellcolor{white} \tiny{planets (compressible planets)}  \\
\hline

\RowColor       \cellcolor{white} & \cellcolor{blue} & \cellcolor{white} & \textcolor{blue}{How do exoplanets evolve?} \cellcolor{white} &  \tiny{Requires multiple elements: planetary compositions, formation and evolution} \cellcolor{white} & \cellcolor{white} \tiny{super-Earths to giant planets,} 
\\ 
\RowColor       \cellcolor{white} & \cellcolor{blue} & \cellcolor{white} &  \cellcolor{white} &  \tiny{models. Observing young planets is of high importance} \cellcolor{white} & \cellcolor{white} \tiny{young planets}  \\
\hline

\RowColor       \cellcolor{red} & \cellcolor{blue} & \cellcolor{white} & What is the inflation mechanism  \cellcolor{white} &  \tiny{Ohmic dissipation is the leading contender. {\it Ariel} can lead to significant progress by characterizing} \cellcolor{white} & \cellcolor{white} \tiny{hot Jupiters \& warm Jupiters}
\\ 
\RowColor       \cellcolor{red} & \cellcolor{blue} & \cellcolor{white} &  \cellcolor{white} of Hot Jupiters? &  \tiny{the atmospheres of several hot Jupiters. Measuring wind speeds and atmospheric compositions.} \cellcolor{white} & \cellcolor{white} \tiny{for comparison}  \\
\hline

\RowColor       \cellcolor{red} & \cellcolor{white} & \cellcolor{green} & What is the source of super metal-  \cellcolor{white} &  \tiny{Characterize the atmospheric composition of super metal-rich (dense) gas giants.} \cellcolor{white} & \cellcolor{white} \tiny{A few selected dense gas giants.} 
\\ 
\RowColor       \cellcolor{red} & \cellcolor{white} & \cellcolor{green} &  \cellcolor{white} rich gas giants? &  \tiny{This will provide information on the atmosphere-interior connection, and their formation path.} \cellcolor{white} & \cellcolor{white}  \\
\hline

\RowColor       \cellcolor{red} & \cellcolor{blue} & \cellcolor{white} & What is the transition from   \cellcolor{white} &  \tiny{Atmospheric compositions of objects as a function of mass } \cellcolor{white} & \cellcolor{white} \tiny{$\sim$10s of targets}
\\ 
\RowColor       \cellcolor{red} & \cellcolor{blue} & \cellcolor{white} &  \cellcolor{white} gas giant to ice giant? &  \tiny{at a mass of $\sim$ 0.3 M$_J$.} \cellcolor{white} & \cellcolor{white}  \\
\hline

\RowColor       \cellcolor{red} & \cellcolor{blue} & \cellcolor{white} & What is the transition from   \cellcolor{white} &  \tiny{Atmospheric compositions of objects as a function of mass } \cellcolor{white} & \cellcolor{white} \tiny{$\sim$ 10s of targets}
\\ 
\RowColor       \cellcolor{red} & \cellcolor{blue} & \cellcolor{white} &  \cellcolor{white} ice giants to super-Earths? &  \tiny{at a mass of a few to $\sim$ 10 M$_{\oplus}$.} \cellcolor{white} & \cellcolor{white}  \\
\hline

\RowColor       \cellcolor{red} & \cellcolor{blue} & \cellcolor{white} & What is the role of atmospheric \cellcolor{white} &  \tiny{Atmospheric composition of close in planets from Hot Jupiters to super-Earths} \cellcolor{white} & \cellcolor{white} \tiny{$\sim$10s of targets (0.3 M$_J$-}
\\ 
\RowColor       \cellcolor{red} & \cellcolor{blue} & \cellcolor{white} &  \cellcolor{white} evaporation? &  \tiny{at a mass of a few $\sim$ 10 M$_{\oplus}$.} \cellcolor{white} & \cellcolor{white} \tiny{10 M$_{\oplus}$) focusing on hot ones.} \\
\hline

\RowColor       \cellcolor{red} & \cellcolor{blue} & \cellcolor{green} &  \cellcolor{white} {What planets retain H-He}  &  \tiny{{Atmospheric spectra (scale height)}} \cellcolor{white} & \cellcolor{white}  \tiny{{super-Earths to sub-Neptunes}}\\

\RowColor       \cellcolor{red} & \cellcolor{blue} & \cellcolor{green} &  \cellcolor{white} {atmospheres?}  &  \tiny{{}} \cellcolor{white} & \cellcolor{white}  \tiny{{}}\\
\hline

\RowColor       \cellcolor{red} & \cellcolor{blue} & \cellcolor{green} & What is the distribution of heavy \cellcolor{white} &  \tiny{Understanding the connection between  atmospheric and bulk metallicity.} \cellcolor{white} & \cellcolor{white} \tiny{$\sim$10s of (warm) giant planets}
\\ 
\RowColor       \cellcolor{red} & \cellcolor{blue} & \cellcolor{green} &  \cellcolor{white} elements in giant planets? &  \tiny{Importance of convective mixing and non-adiabatic interiors.} \cellcolor{white} & \cellcolor{white} \tiny{with different mean densities} \\
\hline

\RowColor       \cellcolor{red} & \cellcolor{blue} & \cellcolor{green} & What is the relation between planet \cellcolor{white} &  \tiny{Atmospheric boundary conditions for a few objects in the Neptune to} \cellcolor{white} & \cellcolor{white} \tiny{$\sim$20+ well-characterized objects}
\\ 
\RowColor       \cellcolor{red} & \cellcolor{blue} & \cellcolor{green} &  \cellcolor{white} metallicity and planet mass?  &  \tiny{the brown dwarf regime with well-determined masses and radii.} \cellcolor{white} & \cellcolor{white}  \\
\hline

\RowColor       \cellcolor{red} & \cellcolor{blue} & \cellcolor{green} & What is the relation between planet \cellcolor{white} &  \tiny{Atmospheric boundary conditions for a few objects in the Neptune to} \cellcolor{white} & \cellcolor{white} \tiny{$\sim$20+ well-characterized objects}
\\ 
\RowColor       \cellcolor{red} & \cellcolor{blue} & \cellcolor{green} &  \cellcolor{white} metallicity and stellar metallicity?  &  \tiny{Jupiter regime with well-determined masses and radii.} \cellcolor{white} & \cellcolor{white}  \\
\hline

\RowColor       \cellcolor{red} & \cellcolor{blue} & \cellcolor{white}  & What is the role of tides \cellcolor{white} &  \tiny{Characteristic properties of the atmospheres of eccentric hot Jupiters.} \cellcolor{white} & \cellcolor{white} \tiny{$\sim$ 10 eccentric hot Jupiters}
\\ 
\RowColor       \cellcolor{red} & \cellcolor{blue} & \cellcolor{white} &  \cellcolor{white} for the evolution of giant planets?  &  \tiny{} \cellcolor{white} & \cellcolor{white}  \\
\hline

\RowColor       \cellcolor{red} & \cellcolor{white} & \cellcolor{green}  & What is the ice to rock ratio \cellcolor{white} &  \tiny{Measure atmospheric abundances of volatiles vs refractories} \cellcolor{white} & \cellcolor{white} \tiny{super-hot planets}
\\ 
\RowColor       \cellcolor{red} & \cellcolor{white} & \cellcolor{green} &  \cellcolor{white} of planets?  &  \tiny{of close-in planets} \cellcolor{white} & \cellcolor{white}  \\
%\hline
%\RowColor       \cellcolor{red} & \cellcolor{blue} & \cellcolor{green} & Are there (iron) coreless earthlike   \cellcolor{white} &  \tiny{Atmospheric spectra (composition)} \cellcolor{white} & \cellcolor{white} \tiny{small planets}
%\\ 
%\RowColor       \cellcolor{red} & \cellcolor{blue} & \cellcolor{green} &  \cellcolor{white} planets ?   &  \tiny{} \cellcolor{white} & \cellcolor{white}  \\
\hline

\RowColor       \cellcolor{red} & \cellcolor{blue} & \cellcolor{green} & How abundant are clouds?    \cellcolor{white}   &  \tiny{Atmospheric spectra (composition)} \cellcolor{white} & \cellcolor{white} \tiny{all planets} \\

\RowColor       \cellcolor{red} & \cellcolor{blue} & \cellcolor{green} &  \cellcolor{white}   &  \tiny{} \cellcolor{white} & \cellcolor{white}  \\
\hline

\RowColor       \cellcolor{red} & \cellcolor{blue} & \cellcolor{green} &  {How many terrestrial planets} \cellcolor{white}   &  \tiny{{Atmospheric spectra (composition)}} \cellcolor{white} & \cellcolor{white} \tiny{{super-Earths}} \\

\RowColor       \cellcolor{red} & \cellcolor{blue} & \cellcolor{green} &  \cellcolor{white} {have significant atmospheres?}  &  \tiny{{Phase curve measurements}} \cellcolor{white} & \cellcolor{white}\tiny{}  \\
\hline

\RowColor       \cellcolor{red} & \cellcolor{blue} & \cellcolor{green} & Are hemispherical differences in tidally \cellcolor{white}   &  \tiny{Atmospheric spectra (composition)} \cellcolor{white} & \cellcolor{white} \tiny{small planets } \\
 
\RowColor       \cellcolor{red} & \cellcolor{blue} & \cellcolor{green} &  \cellcolor{white} locked rocky planets observed?    &  \tiny{} \cellcolor{white} & \cellcolor{white}  \\
\hline

%\RowColor       \cellcolor{red} & \cellcolor{blue} & \cellcolor{green} &  Do rocky planets have a magnetic  \cellcolor{white}   &  \tiny{Atmospheric spectra (composition)} \cellcolor{white} & \cellcolor{white} \tiny{small planets} \\

%\RowColor       \cellcolor{red} & \cellcolor{blue} & \cellcolor{green} &  \cellcolor{white} field at any size?  &  \tiny{} \cellcolor{white} & \cellcolor{white}  \\
%\hline

\RowColor       \cellcolor{red} & \cellcolor{blue} & \cellcolor{green} &  {What is the variety in the surface} \cellcolor{white}   &  \tiny{{Atmospheric spectra (composition)}} \cellcolor{white} & \cellcolor{white} \tiny{{lava planets like 55 Cnc e}} \\

\RowColor       \cellcolor{red} & \cellcolor{blue} & \cellcolor{green} &  \cellcolor{white} {compositions of close-in rocky planets?}  &  \tiny{{(Especially, the redox state and the amount of volatile element)}} \cellcolor{white} & \cellcolor{white}  \\
\hline

%\RowColor       \cellcolor{red} & \cellcolor{blue} & \cellcolor{green} &  {What is the diversity of atmospheres of planets} \cellcolor{white}   &  \tiny{{Atmospheric spectra (composition)}} \cellcolor{white} & \cellcolor{white} \tiny{{all planets}} \\

%\RowColor       \cellcolor{red} & \cellcolor{blue} & \cellcolor{green} &  \cellcolor{white} {around different  vs.~the same star?} &  \tiny{{Atmospheric spectra (composition)}} \cellcolor{white} & \cellcolor{white}\tiny{}  \\
%\hline
%\hline
%\hline

% \begin{table}
% %\centering 
% \begin{flushleft}
% \footnotesize{
% \begin{tabular}{| l  |l | l | l | l | l  |}

\end{tabular}
}
\end{flushleft}
\caption{A table summarising the key questions to be answered with Ariel. The colours correspond to the different aspects of composition (red), evolution (blue), and origin (green). The required measurements and suggested targets are given in the middle and right columns, respectively. }\label{table}
\end{table}
\end{landscape}

\begin{figure}[h!]
%\floatbox[{\capbeside\thisfloatsetup{capbesideposition={left,top},capbesidewidth=4cm}}]{figure}[\FBwidth]
%\vskip -8pt
%\vspace{-0.8cm}
\centering
%{\includegraphics[scale=.56]{slices_JS.pdf}}
{\includegraphics[scale=.58]{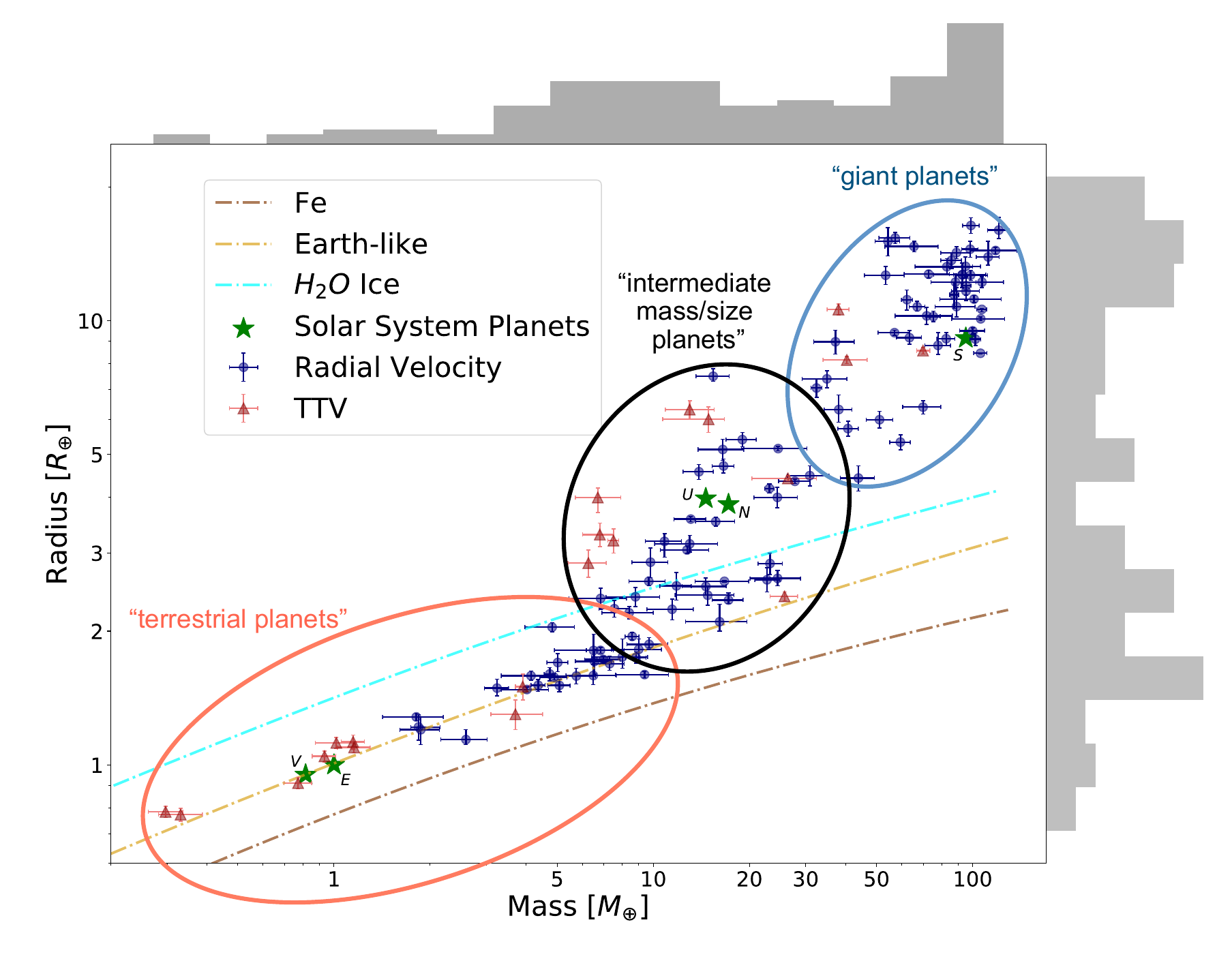}}
{ \caption{\small The M-R diagram of planets with robust mass measurements with relative uncertainties smaller than 25\% for mass and smaller than 8\% for radius. The red triangles and blue circles correspond to data with mass determination from TTVs and RVs, respectively. Also shown are composition lines of pure-iron (brown), Earth-like planets (light-brown) and water ice (blue), and the distribution of exoplanet mass (top) and radius (right). We indicate the planets that are expected to be ``terrestrial'' ``gaseous'' or ``intermediate'' in terms of composition \citep[from][]{2020A&A...634A..43O}.}
\label{fig:MR}}. 
\end{figure}

Figure~\ref{fig:MR} shows the mass-radius (M-R) relation of planets up to a mass of 120 M$_{\oplus}$ as presented by \cite{2020A&A...634A..43O}. 
Identified are the ``terrestrial planets'' whose compositions are dominated by refractory materials, the gaseous (giant) planets that are mostly composed of hydrogen and helium (H-He), and an intermediate population (transitional planets) that includes planets that are massive Earth-like planets (super-Earths) and Neptune-like planets as well as smaller version of the ice giants (mini-Neptunes).

Below we discuss the research conducted related to the questions relevant for {\it Ariel} science in terms of interiors. We organize the discussion based on the masses of the objects as shown in Fig.~\ref{fig:MR}: (i) Giant planets are the planets for which we currently have the most data. (ii) Intermediate-mass planets have slightly smaller radii and are thus not as well characterized, but their high occurrence rate in the galaxy \citep{Fulton+2017} should ensure a large number of potential targets to study with {\it Ariel}. (iii) Super-Earths represent the most challenging class of objects for {\it Ariel}, but they are a milestone towards the characterization of planets more similar to our own Earth. 
%Progress expected from the observation of these classes of planets by 

Our paper is organised as follows. Progress related to gas giant science is discussed in section 2.
Intermediate-mass planets are presented in section 3. 
Science related to terrestrial planets is discussed in section 4.  
{\it Ariel} is discussed in Sections 2 to 4. 
Finally, a summary is presented in section 5. 

\section{Giant Planets}
When it comes to gas giant planets whose compositions are dominated by hydrogen and helium (hereafter, H-He), the planetary bulk metallicity is typically characterised by the metallicity, i.e., the mass fraction of heavy elements within the planet.  
Giant planets are key to understand the formation of planetary system: Dynamically, their migration shaped the final planetary systems, be it our Solar System \citep[e.g.][]{Tsiganis+2005} or exoplanetary systems \citep[e.g.][]{1996Natur.380..606L}. The  gaseous envelopes of giant planets were accreted during the first millions of years of the formation of planetary systems so that the study of their bulk and atmospheric composition informs us on the conditions that led to the formation of planetary systems \citep[e.g.][]{Guillot2005}. 

Because giant planets are H-He dominated
%, it is possible to determine globally the amount of other elements ("heavy elements", i.e., everything that is not hydrogen and helium) that they contain. This is what we term their "bulk" composition. However, contrary to solid planets, giant planets are 
and are compressible they contract as their interior progressively cools \citep[e.g.,][]{Hubbard1977, Guillot+1996, Fortney+2011,vazan13}. This implies that determining their bulk composition from a measurement of mass and radius also requires knowledge of age, equations of state, and atmospheric boundary conditions \citep[e.g.,][]{Guillot1999}. Ariel's observations are thus crucial in the sense that they complement precise determinations of radii of transiting planets and ages of their parent stars, in particular as expected from the {\it Plato} mission \citep{2014ExA....38..249R}.

\subsection{Importance of the atmosphere}
The atmosphere is the external boundary condition used by interior models to calculate the overall structure of a giant planet. It is also a lid that governs how interior heat is progressively radiated away. A proper characterization of the atmosphere is therefore crucial to infer the properties of the planet's interior and its formation. 
For interior models, we seek to obtain with {\it Ariel} several key properties of the atmospheres of giant planets:
\begin{itemize}
    \item Their albedo, which is the main factor governing the equilibrium temperature and therefore atmospheric entropy \citep[e.g.,][]{Guillot+1996}. 
    \item Their day-night and equator-to-pole temperature contrasts, which also govern the rate at which interior heat can leak through \citep[e.g.,][]{Guillot+Showman2002, Rauscher+Showman2014}.
    \item Cloud content and wind properties since these also affect how irradiation energy is absorbed and redistributed in the atmosphere \citep[e.g.,][]{Bodenheimer+2001,Showman+Guillot2002,Batygin+Stevenson2010, Youdin+Mitchell2010, Rauscher+Menou2013}.
    \item Generally, any variation in atmospheric properties, in particular abundances, tracing both the radiative and dynamical properties of the atmosphere \citep[e.g.,][]{Parmentier+2016,Ehrenreich+2020}.
\end{itemize}

\subsection{Inflation of Hot-Jupiters}\label{sec:inflation}
\begin{figure*}
    \centering
    \includegraphics[width = 0.7\columnwidth]{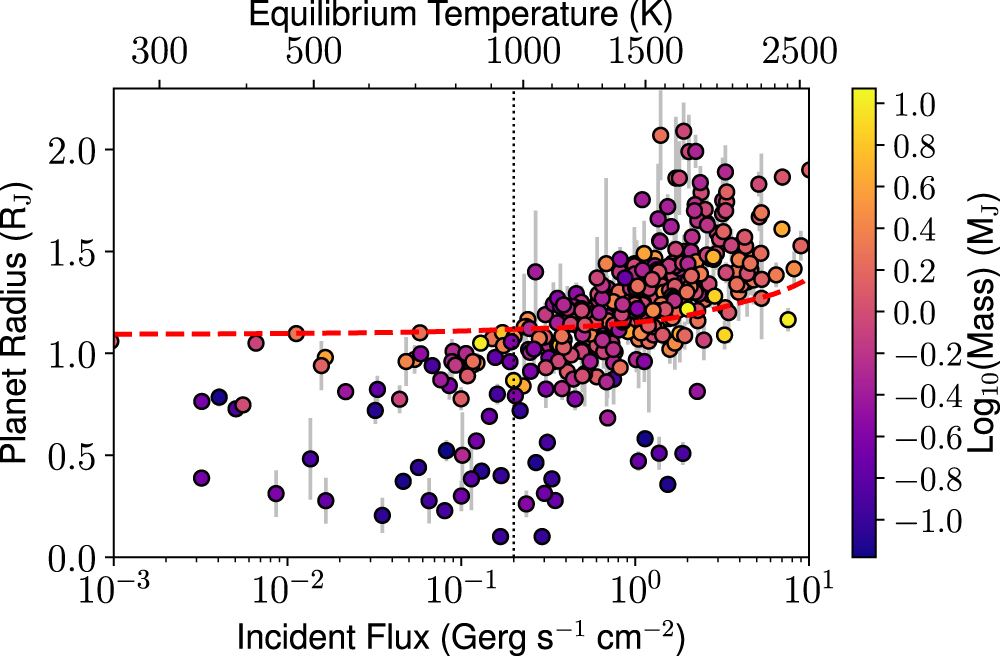}
    \caption{\small Radii of transiting giant exoplanets plotted against their incident flux (or equilibrium temperature) and colored by mass on the log scale. The dashed red line is the radius of a Jupiter-mass pure H/He model with no inflation effect \citep[from][]{Thorngren+Fortney2018}.}
    \label{fig:tf18}
\end{figure*}

The importance of the atmosphere was highlighted by the works of \cite{Bodenheimer+2001} and \cite{Guillot+Showman2002} who showed that the radius of the famous hot Jupiter HD~209458~b was larger than predicted by standard evolution models. 
This inflation of Hot-Jupiters is highlighted with every newly discovered highly irradiated giant planet.  Figure \ref{fig:tf18} demonstrates the clear correlation between the irradiation received by the planet and the decrease in bulk density. Figure \ref{fig:tf18} also  shows that planets with equilibrium temperature above 1200 K have a difference between the observed radius and the one that results from evolution models that increases with $R \propto T_{eq}^{1.4}$ \citep{Laughlin+2011, Miller+Fortney2011, Demory+Seager2011, Laughlin+Lissauer2015, Thorngren+Fortney2018}. 

The inflation of hot Jupiters spurred a series of explanations to explain the observations. These include hydrodynamical dissipation, where the heat gets transported to the interior of the planet through vertical winds that push down kinetic energy that is dissipated into heat \citep{Guillot+Showman2002, Showman+Guillot2002}, heat being transported by turbulent mixing in the external radiative zone \citep{Youdin+Mitchell2010}, or vertical advection of potential temperature by deep atmospheric circulation \citep{Tremblin+2017, SainsburyMartinez+2019}. Another possibility is Ohmic dissipation, resulting from the interaction of the zonal winds with the planets' magnetic field \citep{Chabrier+Baraffe2007, Batygin+Stevenson2010, Perna+2010, Youdin+Mitchell2010, Huang+Cumming2012, Rauscher+Menou2013, Wu+Lithwick2013, Rogers+Showman2014, Ginzburg+Sari2016}. Finally, some studies have remained agnostic towards the physical mechanism that cause the inflation, but showed how different fractions of heat deposition in the planet's interior change its structure and observable parameters \citep{Baraffe+2003, Komacek+Youdin2017}.  

A common denominator from all these studies is that the inflation of hot-Jupiters is caused by the intense stellar irradiation that these planets receive, and therefore changes in the stellar irradiation either due to stellar evolution \citep[e.g.,][]{Ginzburg+Sari2015, Ginzburg+Sari2016, Komacek+2020} or to the planet migration —that changes their semi major axis and the received Flux— \citep[e.g.,][]{Burrows+2000, MolLous+Miguel2020}, can cause a change in the inflation rate of planets, reflecting the planet history. 

{\it Ariel} will aid in in this field with a better determination of composition, thermal profile and circulation in giant planet atmospheres, that will help to find correlations and identify the physical mechanisms that cause hot Jupiter inflation. This is expected to improve our understanding of the internal structure, formation, and evolution of giant exoplanets.

%%%%%%%%%%%%%%% FIGURE %%%%%%%%%%%%%%%
%%%%%%%%%%%%%%% FIGURE %%%%%%%%%%%%%%%

\subsection{Atmospheric and Bulk Composition}\label{sec:composition} 
% Helled

Spectroscopic measurements of giant exoplanets allow us to further characterize these objects. These measurements, combined with accurate mass and radius determinations can be used to better understand the nature of giant planets \cite[see e.g.,][and references therein]{2019ARA&A..57..617M}. 
The bulk composition of giant planets can be estimated from accurate measurements of the mass and radius. This estimate, however, has a relatively large (theoretical) uncertainty since it depends on the materials used to represent the heavy elements, their assumed distribution, and the equations of state used in the models as we discussed below \citep[e.g.,][]{2008A&A...482..315B,vazan13,muller20}. 
Although atmospheric measurements cannot give us the radial distribution of heavy elements directly, it can greatly illuminate this topic when compared to bulk internal enrichment inferred from the mass/radius data alone. 
Indeed, the presence or absence of a systematic bias between the enrichment values given by both methods would allow us to quantify the degree of compositional segregation and inform us on the strength of the mixing processes at play \citep{2013ApJ...775...10V,vazan15}.

The composition of gaseous exoplanets are typically inferred either by assuming a common mechanism for inflated hot Jupiters \citep{Guillot+06,2007ApJ...661..502B} or by analysing only warm giant exoplanets, whose radii are not expected to be affected stellar  irradiation \citep{Thorngren+16}. However, it is not only the bulk composition that is important but the actual distribution of heavy elements within the planet. 
First of all, enriched envelopes have higher molecular weights and therefore shrink more effectively than when the heavy elements are concentrated in a central core. 
This also leads to a higher envelope opacity which leads to a less efficient cooling and slower contraction \citep{Ikoma+06}. This means that at a given age, the estimated heavy-element mass from theoretical models varies depending on the model assumptions. 
This effect is particularly important for planets with significant enrichments and for intermediate-mass planets (sub-giants), since it can result in an overestimate of the heavy-element mass required to reproduce the measured radius \citep{2008A&A...482..315B,vazan13}. 

In order to take full advantage of the expected {\it Ariel} data progress in theory is required. For example, structure and evolution models should consider the uncertainties associated with the assumed opacity, assumed internal structure in particular, the distribution of the heavy elements, the used equation of state, and age of the planet. 
In addition, available opacity tables as a function of metallicity, pressure and temperature are usually sparse and thus, often internal structure models are not self-consistent in terms of the opacity calculation and the envelope metallicity. An attempt to bridge these inconsistencies was presented by \citet{2013ApJ...775...10V}, and a more physically-based opacity model that accounts for various compositions, grain properties and the existence of clouds for gaseous planet is still missing. 

Finally, the connection between the  atmospheric and bulk composition of giant exoplanets needs to be better understood. Even for the solar system gas giants, this relation is still poorly understood and being intensively investigated \citep[e.g.][and references therein]{Helled2019}.

\subsubsection{Compositions of Hot Jupiters}\label{sec:compositionsHJ} 
Current exoplanet data have taught us that there is a large spread in the predicted metallicities of gas giant planets \citep{Thorngren+16}. Therefore there is a clear need to identify the trends in terms of planetary mass--metallicity relation, bulk-composition--atmospheric--composition relation as well as the dependence of the stellar type and age. {\it Ariel} will provide critical information on these relations and will therefore allow us to better characterise exoplanets and will improve our understanding on the dependence of the planetary bulk and atmosphere composition on the stellar and orbital properties.  

%\subsection{Heavy Elements in Gas Giants}
% Formation timescales and lifetime of disks, possibility to capture primary atmospheres, degassing and injection of heavy elements.
%Ikoma 

When the radius, mass, and age of a giant planet are measured by observation, 
the heavy-element content can be inferred through theoretical modelling of the planet's internal structure and gravitational contraction. 
A significant proportion of close-in giant planets detected so far are found to be quite enriched with heavy elements, which account for several tens of percent of the planetary total mass \citep{Guillot+06,Thorngren+16}. 
The first example is HD~149026~b, which is a sub-Jupiter of $\sim$ 110~$M_\oplus$ with inferred metal content of 60--80~$M_\oplus$ \citep{Sato+05}. 
The discovery of such high density giant planets certainly gives support to the core accretion theory of giant planet formation in which a central core composed of heavy elements first forms and then captures the ambient nebular gas composed of H-He in a runaway fashion \citep{Mizuno80,Bodenheimer+86}. 
The standard core-accretion theory does not predict  such high enrichments as inferred for high-density giant planets \citep{Ikoma+06,Helled2014}. 

 One should keep in mind that the inferred composition depends on the material chosen to represent the heavy elements. As a result, the exact composition of giant exoplanets cannot be determined.  
 Indeed, inferring the heavy-element masses in giant exoplanets strongly relies on theoretical modelling. 
It was recently shown by \citet{Muller2020} that the inferred composition of giant exoplanets can significantly vary depending on the model assumptions. Large theoretical uncertainties include the used equation of state, the assumed distribution of the elements, and the atmospheric model. Another important property that should be considered is the planetary age. Its accurate determination can be used to narrow the uncertainty in the inferred composition. We therefore suggest that some of the {\it Ariel} giant planet targets should include planets around stars with a relatively good age measurement (within $\sim$ 10\%). This is expected to be possible thanks to the upcoming {\it Plato} mission \cite{2014ExA....38..249R}. 
We stress that, in order to take full advantage of {\it Ariel} data, progress in theory is required. Although ambiguity on the exact planetary composition are likely to remain, the large statistics expected from {\it Ariel} will be used to identify the trends and improve our understanding of giant planet origins.

\subsubsection{Origin of heavy elements in warm Jupiters}\label{sec:origin_heavies} 
The large amounts of heavy elements in warm Jupiters are expected to be accreted during or after the runaway gas accretion phase. At these later stages the accreted material  does not settle all the way to the center (core) and instead it contaminates the gaseous envelopes. 
According to detailed investigation of the dynamics of planetesimals around a growing proto-gas giant, however, it is hard for a massive planet to capture large amounts of the surrounding planetesimals \textit{in situ} \citep[e.g.,][]{Shibata+19a}. 

\begin{figure*}
    \centering
    \includegraphics[width = 0.49\columnwidth]{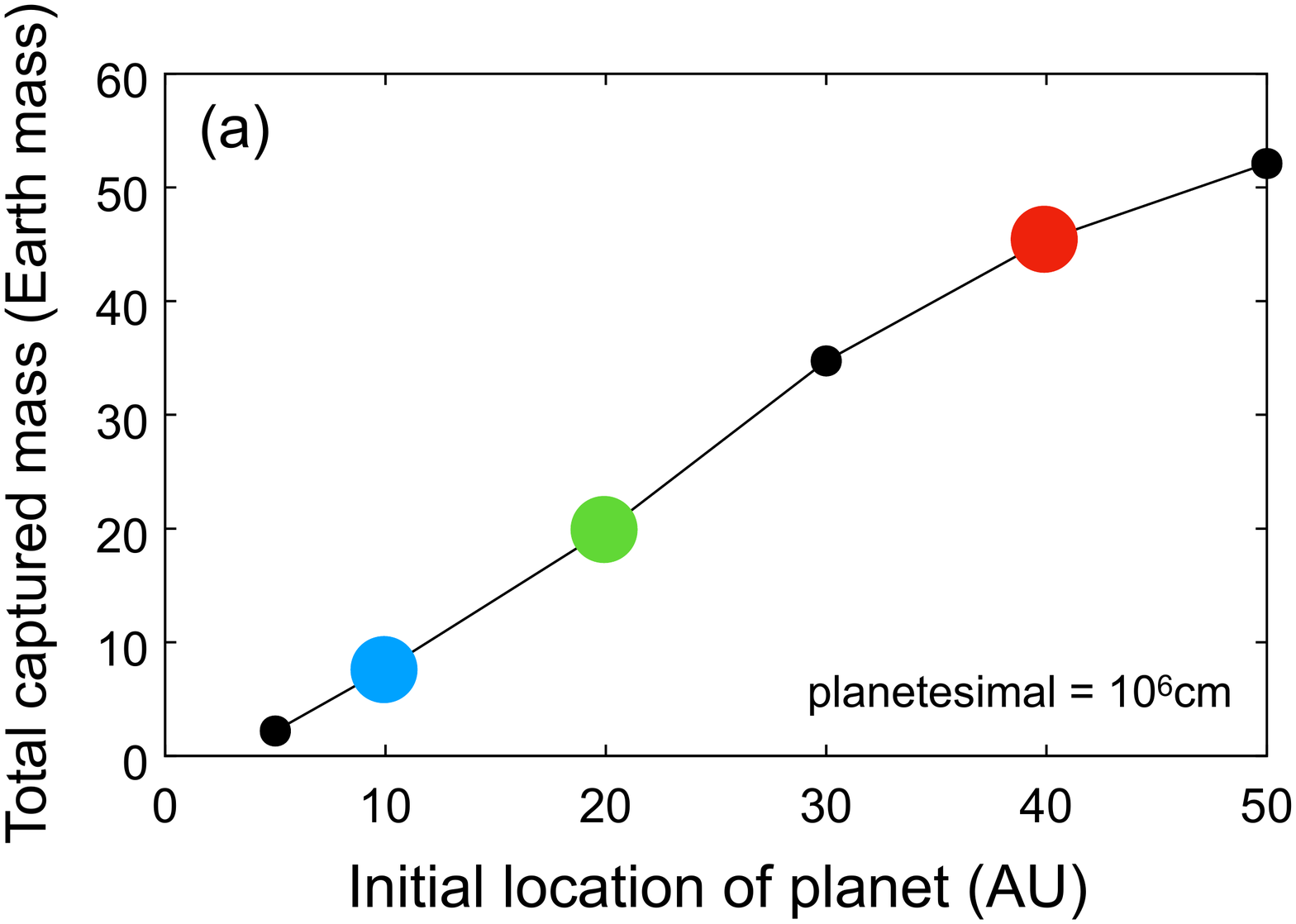}
    \includegraphics[width = 0.49\columnwidth]{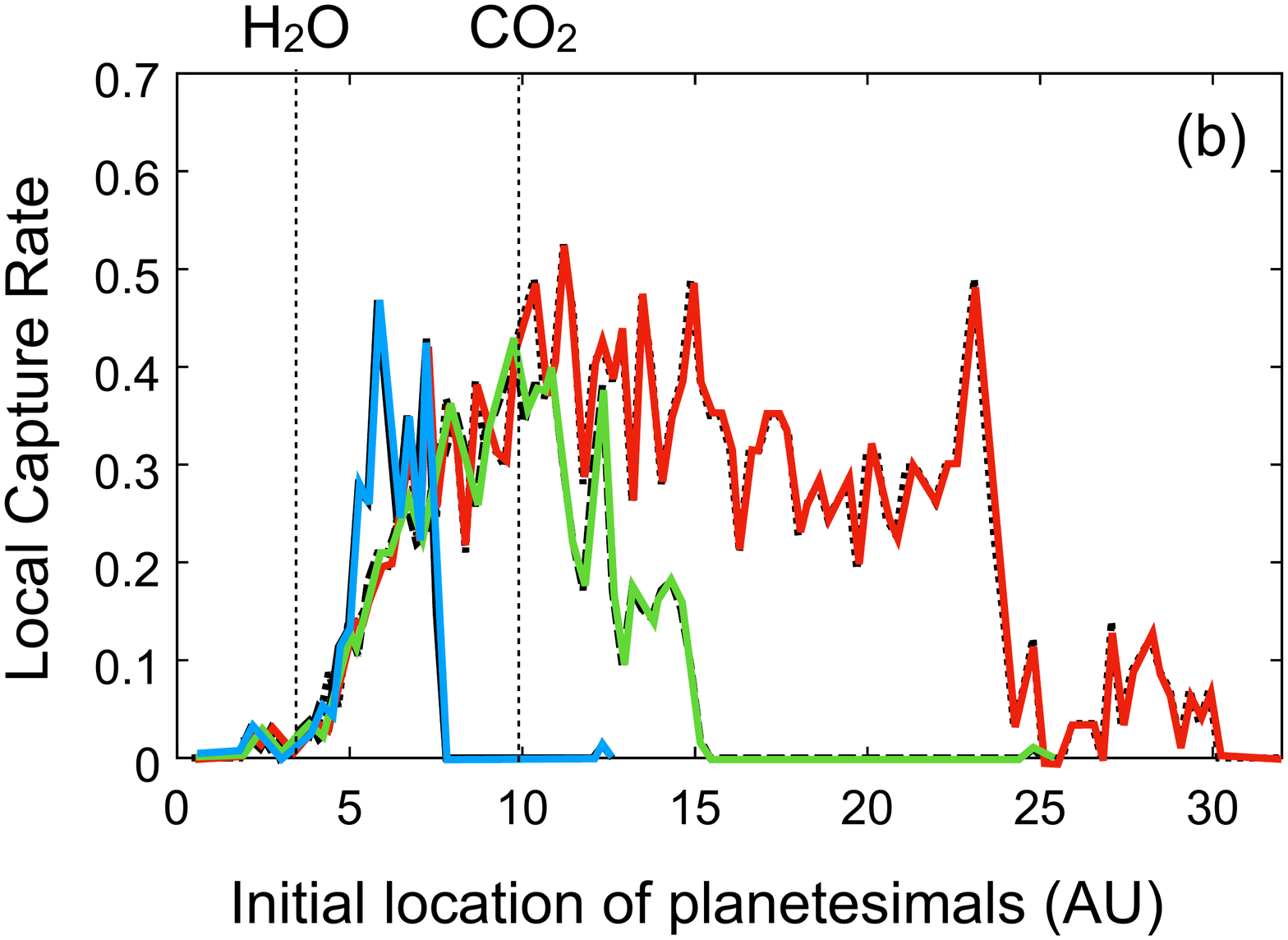}
    \caption{\small Results of dynamical simulations of planetesimal capture by a migrating giant planet---The left panel shows the total mass of the planetesimals that the planet captures during its orbital migration as a function of the semi-major axis at which the planet starts migration; the right panel shows the fraction of the planetesimals located initially at an semi-major axis that the planet engulfs. The vertical dotted lines in panel (b) indicate the semi-major axes of the H$_2$O and CO$_2$ snowlines. Those panels have been adapted from Figure~5 of \citet{Shibata+19b}.}
    \label{fig:Shibata}
\end{figure*}

The effect of planetary inward \textit{migration} on the capture efficiency of planetesimals is shown in Fig.~\ref{fig:Shibata}; the numerical results were obtained by $N$-body simulations for 10-km planetesimals around a migrating Jupiter-mass planet in a protoplanetary gas disc \citep[][]{Shibata+19b}. 
It is confirmed that orbital migration helps the planet capture planetesimals. 
Especially for more than 50-100 $M_\oplus$ of heavy elements to be captured, a long-distance migration ($\gtrsim$~40~AU) is needed, as shown in Fig.~\ref{fig:Shibata}(a).
In addition, planetesimal capture is found to occur in relatively limited regions (see Fig.~\ref{fig:Shibata}[b]); in particular, no planetesimal accretion occurs in inner warm regions due to strong aerodynamic shepherding.
Thus, gas giants migrating over a long distance tend to capture cold materials. 

%%%%%%%%%%%%%%% FIGURE 1 %%%%%%%%%%%%%%%
%%%%%%%%%%%%%%% FIGURE 1 %%%%%%%%%%%%%%%

This leads to a prediction that highly metal-rich gas giants may have not low C/O, but nearly stellar C/O ratios. At such large semi-major axes in passive disks, both carbon and oxygen are contained in ice planetesimals (i.e., beyond the CO$_2$ snowline) from a simple thermodynamic equilibrium consideration \citep[e.g.,][]{Oberg+2011}. Detailed calculations of disk chemistry also demonstrate that the C/O ratio of ice is similar to the stellar ratio at $\gtrsim$~10~au \citep[][]{Eistrup+2016}.  
A determination of the C/O ratio of enriched gaseous exoplanets with {\it Ariel} can be used to test this prediction.

The radii of close-in gas giants are known to increase with decreasing distance to the central stars, on average \citep[see e.g.][]{Spiegel+14}. 
This tendency is not only due to increase in stellar irradiation, but also due to unknown mechanisms for energy injection. 
The latter prevents us from estimating their bulk metallicities correctly. 
Thus, the close-in giant planets with inferred bulk metallicities are ones with moderate irradiation \citep[so-called warm Jupiters; $\lesssim$ a few $\times 10^8$~erg/s/cm$^2$ in][]{Thorngren+16}.
Instead of bulk composition, the atmospheric H$_2$O abundances have been inferred for hot Jupiters with high stellar irradiation via transmission spectroscopy done with Hubble/WFC3 in the near-IR during their primary transits. 
Recent retrieval models for transmission spectra in the optical and near-IR of ten hot Jupiters \citep[e.g.][]{Pinhas+19} show that half of those atmospheres have sub-stellar H$_2$O abundances, though the observational errors are still large. 
This result apparently seems inconsistent with the high bulk metallicities of warm Jupiters. 
However, this is still inconclusive, partly because the argument is based almost only on the water features in the near-IR; and information on  other molecules such as CO, CO$_2$, and CH$_4$ is unavailable. In addition, we do not have enough knowledge of haze and clouds, which possibly obscure the H$_2$O features. 
Finally, the elements and molecules in the atmospheres of giant exoplanets are measured in the uppermost part of the atmosphere and therefore might not represent the bulk composition. Recent developments in giant planet theory suggest that giant planets are likely to be inhomogeneous and have composition gradients \citep[e.g.,][]{2017GeoRL..44.4649W,2017ApJ...840L...4H, Vazan2018, 2019ApJ...872..100D,Ni2019,Helled2019}.  As a result, atmospheric composition only provides a limited glimpse into the composition of giant exoplanets.

A determination of the atmospheric H$_2$O abundance of giant planets is important since it can be linked to their origin and evolution  
\citep[e.g.,][]{2014MNRAS.441.2273H,2017MNRAS.469.4102M}. 
Collecting information on the water abundance in the atmospheres of many hot Jupiters, and comparing them to other elements (e.g., carbon) would reveal important information that could be used to constrain planet formation and evolution models. Nevertheless, connecting this information with the bulk composition remains a challenge.

\subsection{Gradual composition distribution and envelope enrichment by convective-mixing}\label{sec:convective_mixing}
% Vazan
A main challenge in linking the atmospheric composition with the bulk is due to the fact that giant planets might not be homogeneously mixed. This possibility seems to be rather realistic for the solar-system gas giants and there is no reason to believe that giant exoplanets are significantly different. 
At the same time, we are still lacking an understanding of under what conditions giant planets tend to be homogeneous (mass, age, formation process). This topic should be investigated further in order to take full advantage of {\it Ariel} data. 

Composition gradients in giant planets could be a result of number of physical processes such as: (1) Solids (heavy elements) accretion during the formation process \citep[e.g.][]{lozovs17,iarospod07,chattejee18, brouwers18,valleta18,boden18,2020ApJ...900..133V}. (2) Solubility of materials in metallic hydrogen followed by convective mixing  \citep[e.g.][]{stevsalp77,wilson12a,wahl13,soubiran17}. (3) Helium phase separation \citep[e.g.][]{stevsalp77,forthub03,morales09}. 
%(4) Planetesimal dissolution in the gaseous envelope \citep{iarospod07,chattejee18}. 
%(5) Grain sedimentation \citep{mordasini14,ormel14}. 
% Allona, grain settling mostly affect the opacity, not really the internal structure (since the mass of the corresponding layers is rather small), so I rather exclude this point, ok?  Ravit
(4) Rotation and magnetic field effects \citep{chabrier07}. 

The resulting gradual composition distribution can change with time by convective-mixing \citep{vazan15}, which in turn leads to enrichment the outer gaseous layers with deep interior materials. 
The occurrence of convective-mixing depends on the ratio between the temperature gradient and the composition gradient along the interior, according to the Ledoux convection criterion \citep{ledoux47}. 
%For the same mixing parameters the convective-mixing efficiency depends on the mean molecular weight of the metals, as heavier elements are more difficult to mix. Thus, ices are easier to mix by convective-mixing than rocks, as is shown in Fig.~\ref{fig:Vazan1}.

Simulations of self-consistent structure and thermal evolution find convective mixing to be efficient in a large range of giant planets interiors \citep{vazan15,vazan16,muller20}. 
However, since the mixing parameters in planetary conditions are poorly constrained, such simulations provide a range of possible solutions and the efficiency of mixing is yet to be determined. 
For example, mixing could take place in the form of layered-convection, which provides a lower prediction for the efficiency of mixing in giant planet interiors \citep[e.g.,][]{lecontechab12,debraschab19}. 
%a unique solution (see \citet{vazan15} for details). 

Measurements of the atmospheric abundances with {\it Ariel} for large sample of giant planets can help to constrain the parameter-space of the convective-mixing efficiency. 
It is known that planet current location is not necessarily its formation location. In the protoplanetary disk phase young protoplanets are expected to migrate, usually inward \citep[e.g.,][and many more]{tanaka02}, while they are still growing. The gas-to-dust ratio and the chemical element content of different material phases in the protoplanetary disk varies with location and time. We discuss changes with location in the disk, controlled by the temperature. 
% MK: the original sentence was: The available materials in the protoplanetary disk varies with the distance from the star.

Naively, the outer envelope of a gaseous planet is composed of later accreted (current location) materials, while the deep interior composition is related to the planet's formation location.
However, efficient convective mixing sweeps deep interior materials upward, and enriches the outer envelope with formation composition. 
Thus, the abundance of different species by the {\it Ariel} mission, will indicate on the convective mixing efficiency, and on the planet formation location.  

A similar idea was used in several studies to determine Jupiter's formation location, based on its current atmospheric abundances \citep[e.g.,][]{oberg19}. 
Modeling the evolution of Jupiter interior indeed suggests that the early accreted deep interior materials can reach  the outer envelope in less than one giga year \citep{vazan18a}. 
The {\it Ariel} data will allow us to perform similar studies for a large sample of exoplanets, namely to examine material abundances of chemical spices that are not expected to appear in the current planetary location. 
If these elements are found to be abundant the atmospheres, then it would imply that convective mixing from the deep interior is a significant process in giant planet interiors.
Therefore, the {\it Ariel} mission will improve our understanding of convective mixing and material transport processes in giant planet interiors.

\subsection{{\it Ariel} observations of giant planets}
Giant planets are thus key targets to address questions linked to the composition, formation and evolution of planets identified in Table~\ref{table}:
\begin{itemize}
\item The source of the inflation mechanism for hot Jupiters discussed in Section~\ref{sec:inflation} can be better identified by fully characterising the atmospheric properties of both hot and warm Jupiters. Typically, several tens of targets would be needed to improve the statistics and address this topic. 
\item The origin of very metal-rich gas giants that require several tens  M$_\oplus$ of heavy elements is still poorly understood and reflects on their formation and evolution histories. {\it Ariel} observations of a handful of bright, super-enriched gas giants would lead to a robust determination of the planetary radius and atmospheric metallicity to decrease uncertainties in the models. 
\item Understanding the distribution of heavy elements in giant planets requires measuring both their bulk and atmospheric properties. A large-enough statistical sample of tens of targets with well-determined stellar and planetary properties is required.  
\item The role of tides in the evolution of giant planets may be addressed through the characterisation of the atmospheres of highly eccentric planets, in order to get both their atmospheric properties and atmospheric thermal evolution at periapsis. This may be achieved through the characterisation of a handful of bright targets. 
\item The characterisation of atmospheric properties in giant planets, in particular the presence of clouds, their physical and chemical properties and their evolution in time is crucial to understand the planetary long-term evolution and infer bulk compositions. Also in that case, the possibility to observe tens or hundreds of planets with {\it Ariel} to acquire spectra spanning the entire visible to infrared wavelength range is needed.  
%will be a particularly powerful asset.
\item Finally, we stress that the observation of signatures of refractory species like TiO, Fe, Na or S (which may be brought into planets as FeS) may yield constraints on the ice-to-rock ratios in giant exoplanets, a crucial parameter for formation models \citep[see][]{Kunitomo+2018}. 
%that is out of reach of direct measurements in solar system giant planets. 
\end{itemize}

From the viewpoints both of planet formation and planetary interior, of particular interest are the atmospheric compositions of giant planets with estimated bulk compositions. At present, the problem is that different types of samples (warm Jupiters' interiors and hot Jupiters' atmospheres) have been compared. 
There are more than ten target planets that are overlapping between the current {\it Ariel} MRS list \citep{Edwards+19} and \citet{Thorngren+16}'s list, which include
CoRoT-10~b, 
HAT-P-15~b, HAT-P-17~b, HAT-P-20~b, HAT-P-54~b, HATS-6~b, HATS-17~b, 
HD 17156~b, HD 80606~b, 
Kepler-16~b,  
WASP-8~b, WASP-80~b, WASP-84~b, WASP-130~b, and WASP-132~b.
We suggest that these planets are particularly  interesting for understanding the origin of heavy elements of close-in gas giants and also the partitioning of heavy elements between the atmosphere and interior. 

\section{Intermediate-mass Planets}

\subsection{An abundant yet poorly known class of planets}

\begin{figure}
    \centering
    \includegraphics[scale=.22]{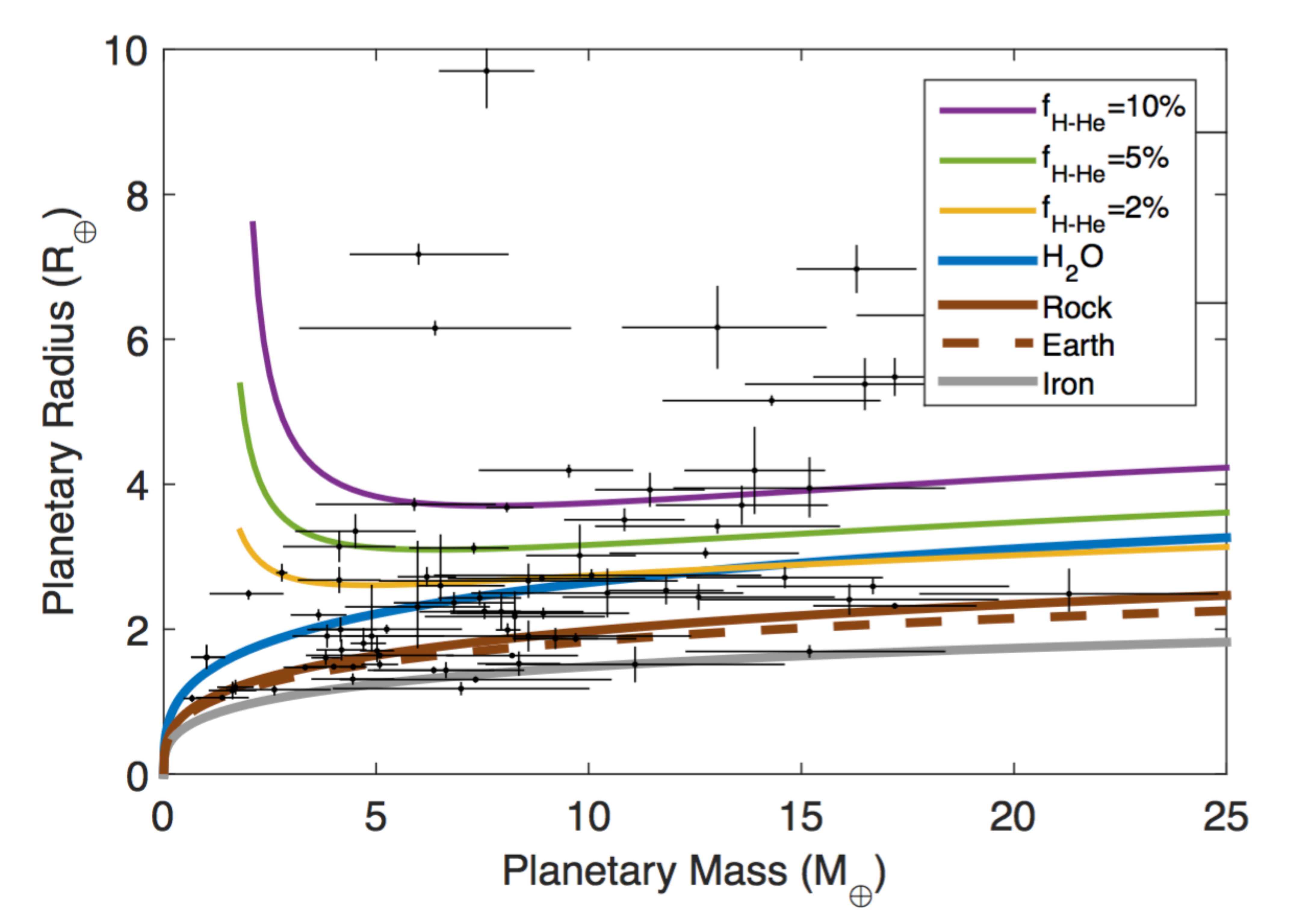}
    \caption{Theoretical mass vs.~radius relationships for \bf{planets of pure iron, rock (MgSiO$_3$) and water composition as well as Earth-like interior, in addition to rocky planets harboring} H-He atmospheres with mass fractions of 2\%, 5\% and 10\%. The black dots with the error-bars show a sample of small and intermediate mass exoplanets.  The figure is taken from \citet{lozovsky2018threshold}.}
    \label{fig:my_label}
\end{figure}
% C.D.: I think it would be good to show also M-R data of confirmed exoplanets in Fig.4 and to state for which irradiation the M-R curves are calculated.

While in the Solar System there is a clear division between terrestrial and giant planets by mass and/or size, exoplanet data have taught us that although planets can be refractory or H-He dominated, there is a non-negligible population of planets that have intermediate masses (1-20 $M_{\oplus}$) and radii (1-4 $R_{\oplus}$), \citep{2020A&A...634A..43O}. 
These planets represent a unique planetary class -- it is not possible to simply re-scale models of the terrestrial or gas giant planets. 
Intermediate-mass planets can be larger versions of terrestrial planets (e.g., super-Earths) or smaller versions of giant planets (e.g., mini-Neptunes) but could also be a class of planets that have different compositions than what is typically assumed, such as iron-coreless \citep{Elkins-Tanton+2008}, carbon-rich \citep{Madhusudhan+2012, Miozzi+2018}, water-rich \citep{2003ApJ...596L.105K,2004Icar..169..499L}, or Ca-Al-rich \citep{dorn2019new} planets.
The transition between terrestrial-like to gaseous-like (H-He dominated) planets is unclear, and at a given planetary mass a planet can belong to either of these populations \citep{2008ApJ...673.1160A}.  This is demonstrated in Figure 1, where we show the M-R relation of exoplanets with well-determined masses and radii. 

This intermediate "overlapping" population is of great interest to the planetary community since the formation and evolutionary paths of such planets are poorly understood. This is true both for super-Earths \citep{2019A&A...627A..83L} and mini-Neptunes  \citep{2014ApJ...789...69H,2017ApJ...848...95V}. It may be possible to distinguish between dominantly rocky, icy or gaseous planets, when additional information on the age of the exoplanetary system exists and the composition of the planetary atmosphere becomes known. In Figure 4 one can see the intermediate-mass/size planetary population. 
Several key questions linked to intermediate-mass exoplanets include: 
\begin{itemize}
\item[$\bullet$] What is the atmospheric composition of  intermediate-mass planets?
 \item[$\bullet$] Are the atmospheres of intermediate-mass exoplanets primordial?
  \item[$\bullet$] What are the typical compositions of  intermediate-mass planets?
   \item[$\bullet$] Are intermediate-mass planets rich in water?
 \item[$\bullet$] How do the atmospheres of intermediate-mass planets interact with their deep interiors/surfaces?
\end{itemize} 

Intermediate-mass planets of several Earth masses with gaseous envelopes, are expected to stay in a molten phase for giga-years \citep{vazan18b}. The long-last interaction of the molten magma-ocean surface with the convective envelope can enrich the envelope with metals, and affect the planet atmospheric abundances \citep{kite19}. Yet, the partitioning behaviour of volatiles into magma at increasingly high pressures is uncertain and requires concerted experimental effort to connect atmospheric models of extrasolar planets to petrologic data. 
Measurements of the atmospheric abundance of magma-ocean planets with gas envelopes by the {\it Ariel} mission will put constraints on the efficiency of the magma-envelope interaction. 

Water-rich interiors are also possible scenarios for intermediate-mass planets of radii below 2.6 $R_{\oplus}$ \citep{lozovsky2018threshold}. For example, stars of very low-mass or low content of $^{26}$Al are thought to host planets rich in water \citep{Lichtenberg+19}. However, mass and radius alone cannot distinguish between gas-rich or water-rich interiors and additional constraints on atmospheric composition by {\it Ariel} are key to further reduce the degeneracy.

 In the Solar System, this class of planets is only represented by Uranus and Neptune, which are poorly understood \citep[e.g.,][]{Helled+2020, HelledF2020}. Observations of planets with similar masses and radii of Uranus and Neptune with {\it Ariel} would provide highly informative and complementary statistical information on a wide variety of planets with intermediate masses/sizes. A future mission to Uranus and Neptune \citep[e.g.,][]{Guillot+2020, Fletcher+2020} would give us the keys to really understand Uranus and Neptune and reflect this knowledge on intermediate-mass exoplanets.

\subsection{{\it Ariel} observations of intermediate-mass planets}
Like gas giant planets, intermediate-mass planets are key targets to address questions linked to the composition, formation, and evolution of planets as identified in Table~\ref{table}. 
%Their smaller size is compensated by a higher frequency in the Galactic population \citep[e.g.,][]{Fulton+2017}, 
Given their common occurrence in the galaxy \citep[e.g.,][]{Fulton+2017}, we can expect significant improvement in the characterization of this diverse and still mysterious class of planets with {\it Ariel}. 
In particular, key observations include: 
\begin{itemize}
    \item The transition from gas to ice giants needs to be well understood, in particular by characterizing atmospheric compositions of a variety of planets with masses $\sim 0.3\rm\,M_J$. Given the large parameter space (in terms of orbital period, composition, eccentricity, stellar properties), we envision that a large ensemble of tens or hundreds of planetary atmospheres should be characterised by {\it Ariel}. This will lead to a  significant progress in our understanding of the nature of intermediate-mass planets and their formation mechanism. 
    %of planetary systems in general.  
    \item Because of their smaller mass, and therefore, lower gravity and limited atmospheric/envelope mass, intermediate-mass planets, together with  Super-Earths, are crucial targets to  understand atmospheric evaporation. 
    %Since atmospheric loss affects the evolution of planets close to their stars, which are more easy to characterised in the near future, it is an important to understand it and a natural outcome of the expected Ariel observations. 
    Observations of tens to hundreds of planets with intermediate masses that orbit close to their stars class are  desirable. 
    \item Intermediate-mass planets should show a much larger variety of atmospheric compositions than giant planets, owing to the potentially low abundance of hydrogen and helium (or conversely, the potentially high metallicity) in their atmospheres. Studies of their atmospheric compositions can be used to understand the link between ice-to-rock ratio and planet formation mechanisms, the role of clouds in planetary atmospheres, and the relations between atmospheric and bulk composition. 
\end{itemize}

\section{Terrestrial Planets}

Recent studies indicate rocky exoplanets are common \citep{2019AREPS..47..141J}. 
So far, over 1000 exoplanets whose radii are less than 2 $R_\oplus$ have been discovered. At the lower radius end of these size range (1-2 $R_{\oplus}$) the planet classification enters the terrestrial planet regime \citep{Owen+2017,jinmordasini2018}. \citet{Fulton+2017} suggests that there is a detection deficiency around 1.5-2 $R_\oplus$ for small close-in planets so that the observed limit may rather be 1.5 $R_\oplus$. 

Observations of terrestrial planets are generally restricted to radius and mass, thus the bulk density can be determined. Detailed interior structure models of terrestrial planets  are obtained by combining mineral physics and average composition. Predictions for Earth-like exoplanets derived from solar system terrestrial planets are limited by the defined composition, which results in iron cores, silicate mantles and negligible but visible atmospheric mass. 

During the main epoch of accretion, rocky planets likely melt entirely due to release of potential energy \citep{2012AREPS..40..113E}, of short-lived radioactive isotopes as in the Solar System \citep{2006M&PS...41...95H,2016Icar..274..350L}, and thermal blanketing of the captured nebular proto-atmosphere \citep{Ikoma+18,2019PEPI..29406294O}. As a result Fe metal, which is immiscible with silicate, is able to sink to the planet's centre forming a metal core (carrying with it elements with an affinity to chemically bind to Fe, such as Ni and limited amounts of light elements, e.g., H, C, N, O, S, Si, \cite{2016AmMin.101..540H,2019PNAS..11614485D,Fischer8743,2019GeoRL..46.5190H}). The silicate mantle left behind undergoes further differentiation, producing a crust and atmosphere. Such differentiated planets may enter a range of geodynamic regimes, of which Earth's plate tectonics is one example \citep{2018RSPTA.37680109S}.  The geodynamic regime entered is intimately linked with a planet's thermal history, influencing whether a magnetic field develops due to core convection, the structure and stability of planetary crusts, rates of volcanism, and the efficiency of surface recycling \citep{foleydriscoll2016}. A comprehensive review of the solar system terrestrial bodies can be found in \citet{tronnes+2019}. 

\begin{figure}[h]
\centerline{\includegraphics[scale=.85]{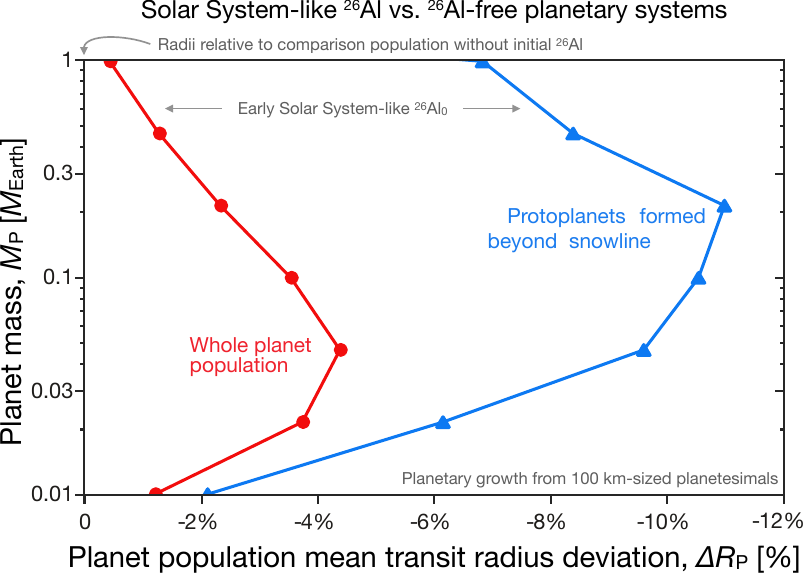}}
\caption{ 
\small{
Predicted deviations in $M$-$R$ diagram for rocky planets formed with solar system-like abundances of the short-lived radionuclide $^\mathrm{26}$Al compared to $^\mathrm{26}$Al-free systems. Planets with similar or higher than Solar abundances tend to form dried-out rocky planets; 
planets in $^\mathrm{26}$Al-free systems are statistically significantly enriched in water. The abundances of $^\mathrm{26}$Al scale with the mass of the star-forming environment of a given planetary system \citep{Lichtenberg+16b}, suggesting a significant divergence 
in rocky planetary radii across these two regimes: planetary systems from massive star-forming regions form terrestrial planets; those from low-mass environments statistically tend to form ocean worlds with water contents on the order of $\sim$10\%. Figure modified from \citep{Lichtenberg+19}.
}
}
\label{fig:TL2}
\end{figure}

The history of Earth, Venus, and Mars demonstrates the diversity of terrestrial planet atmospheres. The mass and composition of an atmosphere of terrestrial planets evolves through delivery of volatiles by nebular ingassing \citep{2019Natur.565...78W,2019PEPI..29406294O}, volatile-ice rich precursors \citep{2015Icar..248...89R}, outgassing from the rocky interior \citep{hirschmann2012,2017ApJ...843..120S,Ikoma+18}, and loss to space \citep{Ikoma+18}. Delivery by solid phases due to planetesimals/pebbles is expected to dominate during the early stages, and may be altered due to their internal geophysical evolution \citep{Lichtenberg+19} that can alter the structural properties of rocky planets in a statistical fashion (Fig. \ref{fig:TL2}). Outgassing occurs during magma ocean cooling \citep{hirschmann2012,2016ApJ...829...63S,2017JGRE..122.1458S,Bower+18,Ikoma+18} but can continue during a planet's life-time through volcanism (e.g., like we know it from Earth today). 

The chemistry and efficiency of outgassing/ingassing are controlled by many aspects among which some are largely unconstrained (e.g., mantle dynamics, thermal state) and some can be constrained (e.g., surface gravity, bulk composition, redox state).  While small terrestrial  exoplanets cannot be probed directly, their composition and evolution may be inferred from knowledge of the thickness and composition of their atmospheres as constrained by {\it Ariel} observations. However, further progress in theory is required to take full advantage of {\it Ariel} data.

Like for the gaseous planets, major challenges for understanding the planetary interior structure relate to our progress in developing structure and evolution models. 
We need to further investigate uncertainties associated with the assumed interior structure, which are the abundance of elements, the used equations of state, and the actual size of the planet, as well as the effect of the stellar radiation on the planet. 
Observations by {\it Ariel} could permit inferences and allow for constraining or discarding current model postulates, including several aspects such as implications for interior dynamics, crusts, and atmospheres. 

\subsection{Bulk composition and consequences for secondary atmospheres}

Planets that form within the same proto-planetary disk can have very different volatile element budgets \citep[e.g.,][]{oberg2016excess} but are expected to have similar budgets in relative refractory elements \citep[e.g.,][]{elser2012origin}. The reason is that condensation fronts of refractory compounds (e.g., of Al,
Ca, Mg, Si, Fe, Na) occur within a small region near the star, whereas condensation fronts of volatile compounds (incl. S, C, O, N, He, H) occur in a very extended region around the star \citep[e.g.,][]{wang2019volatility}. Chemical kinetics timescales for volatile elements also become comparable to other evolutionary timescales in the disk, adding variation to the fraction available as condensates or vapor.  Therefore, relative abundances of refractory and rock-forming elements have been thought to be, to first order, identical between the host star and the planetary building blocks and eventually the planets. 
As a result, it is often assumed that Mg/Si and Fe/Si ratios of the planet bulk can be directly informed by the host star  abundance \citep{dorn+2015,unterborn2017effects}. The majority of planet hosting stars have molar Mg/Si-ratios between 0.7 and 1.5, and molar Fe/Si-ratios between 0.5 and 1.0. Within this range, the Solar composition is average in terms of Mg/Si, but near the higher end of Fe/Si. 

However, a recent study suggests that the refractory ratios of both Fe/Si and Fe/Mg ratios have a wider distribution in super-Earth planets compared to that of planet-hosting stars \citep{plotnykov2020chemical}. This finding challenges the assumption that super-Earths and the cores of mini-Neptunes can be assumed to have the same refractory composition as that of the hosting star.
 A recent study by \citet{schulze2020probability} investigates the eleven individual systems with well-characterized Super-Earths for which host star abundances are available. It was found that only one planet does not reflect its host star abundance and is expected to have a Super-Mercury interior.
 
%Adibekyan et al.(in review) further analysed abundances of 21 stars and find strong evidence for a correlation to the composition of their hosted Super-Earths.}

\citet{plotnykov2020chemical} showed that the uncompressed density is an ideal  metric to compare the planetary composition. This is different than the commonly used bulk composition that depends on both pressure-temperature regime and composition. They obtain the uncompressed density of all exoplanets that are consistent with a rocky composition (below the  threshold radius for rocky planets \citep{lozovs17}) and with mass and radius uncertainty less than 25\%. It appears there is a maximum enrichment in iron corresponding to an uncompressed density of $\sim 6$ g cm$^{-3}$.

Fortunately, the {\it Ariel} mission will provide reliable and homogeneous stellar abundances (Danielski et al. in prep., {\it Ariel} Stellar Characterisation WG). In combination with  internal structure models we will be able to test the primordial origin hypothesis for solid planets, determine if there is indeed a maximum iron enrichment possible from formation, and put key constraints on the bulk abundance of terrestrial exoplanets.

The bulk rock composition of planets also has direct consequences on the planetary evolution of and the secondary atmosphere. 
The planetary bulk composition influences tectonic processes that allow volatiles to ingas from the atmosphere to the mantle \citep{unterborn2017stellar} and outgassing through volcanism \citep{dorn2018outgassing}. Furthermore, the bulk rock composition influences whether exoplanet mantles convect in a single layer or experience double-layered convection \citep{spaargaren2020influence}. This has direct implications for the exchange of volatile between reservoirs. 
A comprehensive study that rigorously investigates the effects of bulk rock composition on melting and outgassing is still lacking and further theoretical efforts are needed to better understand the link between bulk rock composition and atmosphere evolution and chemistry.

Finally, a crucial aspect that determines the chemistry of a secondary atmosphere is the mantle oxidation state. Under reducing conditions, the outgassing of H$_2$ and CO is favoured, while oxidising conditions favour H$_2$O and CO$_2$ to outgas \citep{hirschmann2012,2020NatCo..11.2007D}. Changing how reducing an atmosphere is has important implications for prebiotic chemistry \citep{rimmer2019hydrogen} and climate \citep{wordsworth2013hydrogen}. 
%Reducing atmospheres will also be more extended compared to oxidising atmospheres, as their lower mean molecular weights lead to large scale-heights. [..] 
%CD MORE TO ADD HER
% CD: ADD TARGETS HD21, 55 CNC E, ...

\subsection{Interior structure and dynamics} 

%In the solar system, the compositions of the terrestrial planetary bodies result from the compositions of solids obtained from a combination of solar element abundance and distance from the sun. The density of a planetary body (accounting for compression) is then the main clue to its composition. This approach is adopted for evaluating exoplanetary composition as well, while actual observations are restricted to radius and lower mass limit, and thus only the mean density of the planet can be determined. 
The system's redox state of terrestrial planets is of high importance. This is because it determines the core-mantle fraction, as well as the secondary atmosphere composition, which can be dominated either by H$_2$O and CO$_2$ or H$_2$ and CH$_4$ for example. The availability of oxygen in the exoplanetary system and during planetary accretion, allows for core formation due to the oxidation of the mantle \citep[e.g.,][]{frostmccammon2008}. 

%A comprehensive review of the solar system terrestrial bodies can be found in \citet{tronnes+2019}, and shows that in principle, the terrestrial planet interior structure is the same, but the radial proportion of core and mantle is different (Mercury 0.68, Earth 0.32, Mars 0.2), in agreement with an increase in density for decreasing distance to the star \citep[e.g.,][]{jinmordasini2018}.

Super-Earths cover a mass range of up to 10 $M_{\oplus}$, although early studies speculate interior structures with Earth-like mantle-core fraction to up to 20 $M_{\oplus}$, \citep{howe+2014,2020A&A...634A..43O}. 
Given the solar system planet composition range, there is likely an upper limit of super-Earth radius for an Earth-like mean density \citep{kaltenegger2017}, which is at most 2 $R_{\oplus}$ \citep{lozovsky2018threshold}. 
Scaling Earth’s structure to larger objects, \citet{tackley+2013} investigated the potential range of interior dynamics and surface tectonic expressions (stagnant, episodic, mobile), finding that convection still takes place in the interior of large super-Earths but that it can be sluggish, and confirming earlier findings \citep[e.g.,][]{valencia+2007} that larger planets are more likely to display plate tectonics. With possibly only slight variations in the parameter set, however, it has also been found that size may not be important or increased size rather hinder the development of plate tectonics \citep[][and references therein]{foleydriscoll2016}. Conditions can vary due to the variations in the different systems, which include surface temperature (regulated by atmosphere and distance to the central star), internal heating (related to radiogenic elements and/or tidal dissipation), differences in yield stress (due to composition, particularly variations in water content of the rock crystal structure). 

Several studies provide a phase diagram that suggests a complex relation of these parameters for the evolution of surface tectonic regimes\citep[e.g.,][]{oneill+2016}; however, such studies neglected magmatism and crustal production, which can have a first order effect on the tectonic regime, particularly during early, hot phases \citep{lourenco+2018, lourenco+2016}. Thus, there are not only limitations in the numerical implementation of geological processes, but actual lack of knowledge on for example material behaviour under high pressures, when the mixtures are more complex, which hinders more reliable determination of interior structure, composition and evolution of the terrestrial exoplanets.

\begin{figure}
\centering
\includegraphics[scale=.4]{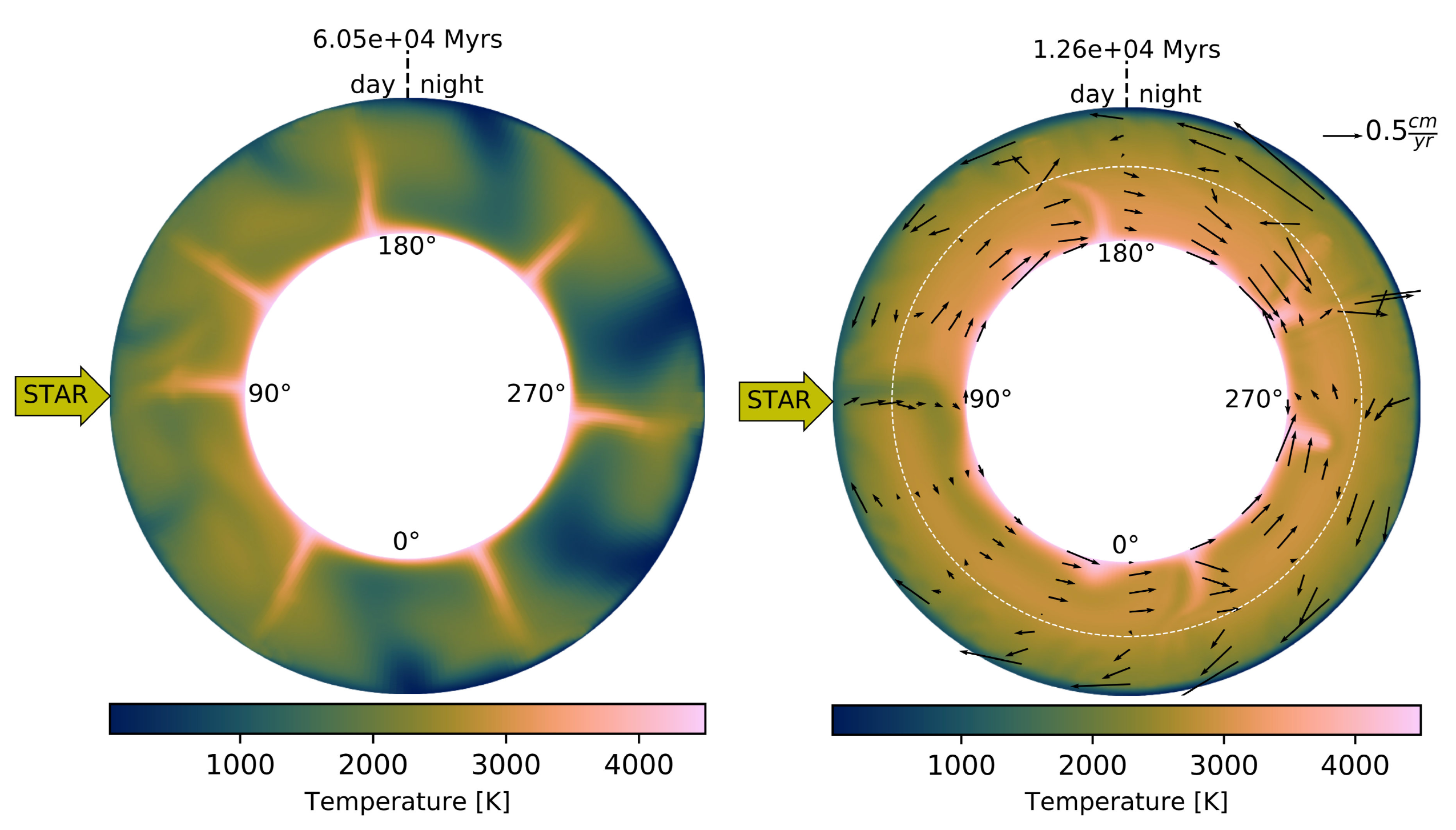}
\label{fig:TL1}
\caption{\small Model of the geodynamic and tectonic state of the interior of LHS 3844b, assuming isothermal surface boundary conditions  according to the phase maps from \citet{Kreidberg+19}. The figure shows the temperature field with a weak (left) and a strong (right) surface layer, with a substellar surface temperature of 1000 K and antistellar surface temperature of 20 K. The distribution of plumes in the hemispheric tectonics regime (right) beneath the substellar point and the distribution of largely molten regions points to a tectonic regime not observed in the Solar System and may provide vital clues as to the potential variability of tectonic states across the rocky exoplanet census. Figure adapted from \citet{2020EPSCMeier}.}
\end{figure}

The orbital setting of close-in planets suggests that they are tidally locked, so that interior dynamics and surface tectonics may differ between the star facing and its opposite side. Several studies investigated the stellar side melting of surface materials and the formation of hemispherical magma oceans \citep{vansummeren+2011}. This molten surface adds a fourth tectonic regime besides, stagnant, episodic or mobile lid known from our solar system. Given the potential of an hemispherical protracted magma ocean, exoplanet interior dynamics may not follow any of the solar system regimes. Observations and phase maps of rocky super-Earths can improve our understanding of the physical mechanisms driving tectonic changes on rocky planets. Furthermore, different regimes of interior-atmosphere exchange relative to Earth and the solar system terrestrial planets will guide our development of more robust geophysical models of the thermo-chemical evolution of rocky proto-planets and their emerging atmospheres. Using constraints from such phase maps \citep[e.g.,][]{Demory+16,Kreidberg+19}, interior dynamical modelling provide insight to interior states of observed super-Earths in order to relate the tectonic, degassing, and atmospheric evolution (Fig. \ref{fig:TL1}). Phenomena related to these hemispherical tectonic regimes may, however, be short-lived, because the planet may continuously reorient due to true polar wander \citep{leconte2018}, and then such a hemispherical difference may be small. 

%%%%%%%%%%%%%%%%%FIGURE%%%%%%%%%%%%%%%%
%%%%%%%%%%%%%%%%%FIGURE%%%%%%%%%%%%%%%%

\subsection{Time dependence of the interior state and implications for atmospheres}

The potential for a terrestrial planet to form (and hold) an atmosphere depends on is mass, as well as the distance to the host star \citep{jinmordasini2018}. The atmospheric composition  depends on the evolutionary stage (age) of the planetary system and the interior structure and interior dynamics generating a magnetic field for the individual planet \citep{boujibar+2020,gaidos+2010}. 

Earth's atmosphere has been profoundly shaped by the presence of life (e.g., the release of oxygen by photosynthesising plants, and sequestration of atmospheric carbon into rocks). The detection of water in the atmosphere may be a requirement for life, but it is insufficient to demonstrate life being present in other systems. At the earliest stages of planet formation, potentially primordial (H-He) composition dominates the atmospheres, but water may be present \citep{Ikoma+18}. Water or steam atmospheres could cause protracted magma oceans and delay cooling of the surface \citep{Bower+18,2019A&A...621A.125B}, prohibiting the formation of lids (lithosphere).

The extent and mass of an atmosphere has a significant impact on the proportions of the solid and gaseous parts of terrestrial planets, and thus on the observed density. The evolution of terrestrial planet atmospheres and speciation of volcanic gases is temperature dependent \citep{hirschmann2012,frostmccammon2008}. Degassing due to volcanism enriches the atmosphere in water, ammonia, methane and/or carbon dioxide, while the primordial atmosphere is lost to space, if the planet is too small, too hot (interior or stellar insulation), or does not have a protecting magnetic field. 
The latter requires a convecting metal core (terrestrial planets) or, in instances where these objects are similar to giant planet’s satellites, subsurface saline oceans. 
The persistence of a magnetic field of the small icy satellites in the solar system requires external heat sources such as tidal interaction with the planet. The persistence of a strong magnetic field of rocky planets is often linked to active plate tectonics. High or low surface temperatures, or climate, has been suggested an important boundary condition of whether plate tectonics would occur. In turn plate tectonic processes have been suggested to moderate climate to be temperate, and allow for cooling of the interior to sustain mantle and core convection to generate the magnetic field.

There are large uncertainties regarding the initiation and sustainability of plate tectonics \citep[][and references therein]{foleydriscoll2016}, which relate to the unknown composition and rheologic properties of the material and the planet, and it remains unclear whether there is a size dependency on the propensity to plate tectonics and similarly the magnetic field strength.

Close-in terrestrial planet are expected to lose their atmosphere due to the stellar radiation pressure. However, if atmospheric gases would be detected by {\it Ariel}, a dominance of H$_2$ and CH$_4$ suggest active degassing, while the dominance of H$_2$O and CO$_2$ may indicate the presence of an atmosphere protecting strong magnetic field, including possible plate tectonics in old systems.

\subsection{Constraints on magma composition of hot rocky super-Earths from atmospheric measurements} 

\begin{figure}[h]
\centerline{\includegraphics[scale=.3]{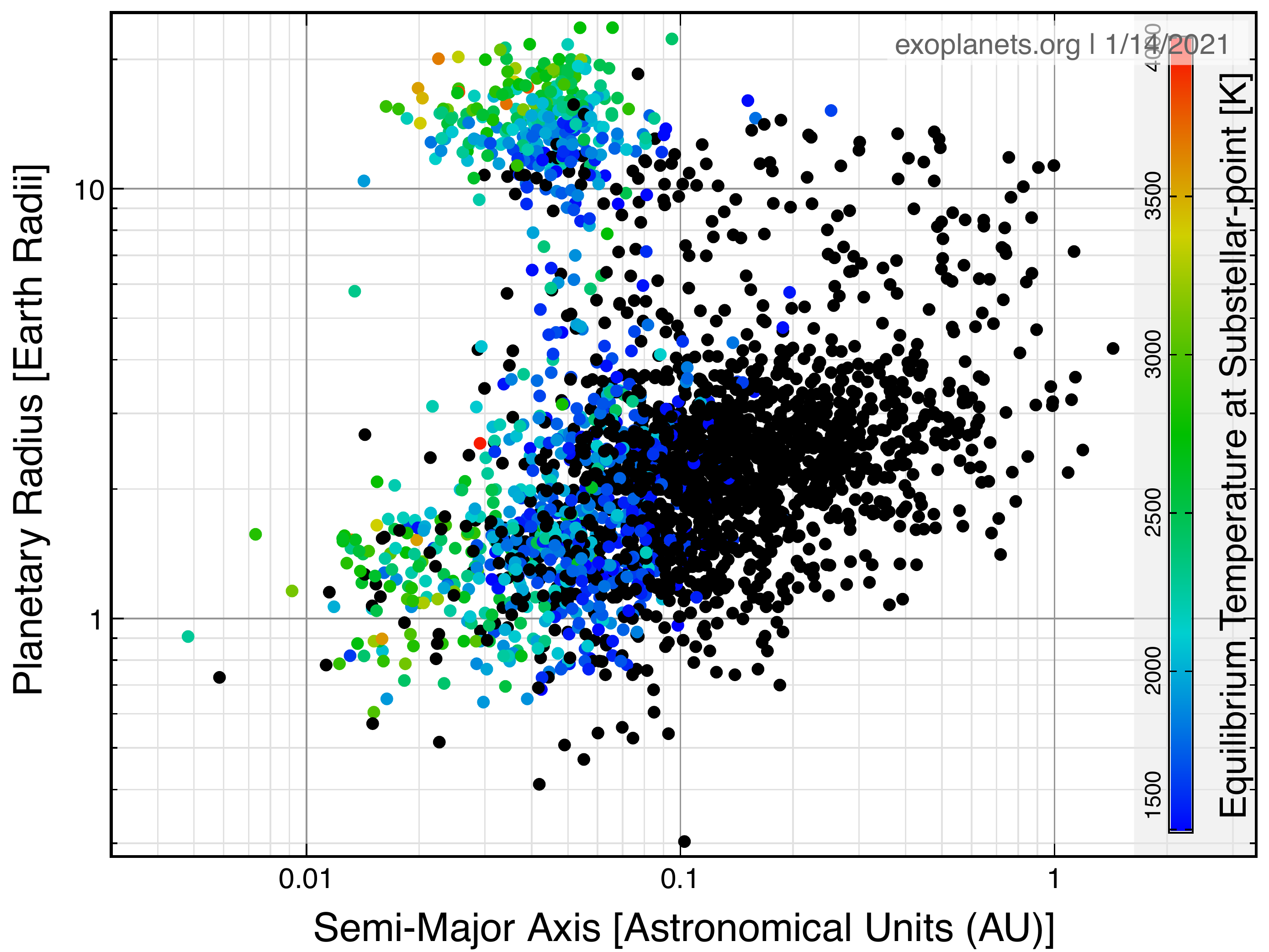}}
\caption{Two-dimensional distribution of discovered exoplanet size and orbit.  The colour contours shows the substellar-point equilibrium temperatures with zero planetary albedo larger than 1500~K. The data has been taken from an open exoplanet catalogue database (http://exoplanets.org)}
\label{fig:dist_exo}
\end{figure}

Over 1000 exoplanets with radii smaller than 2 $R_\oplus$ have been discovered. About half of these planets have substellar-point equilibrium temperatures $T_\mathrm{irr}$ high enough ($\gtrsim$~1500~K) for rock to melt and vaporise (see Fig.~\ref{fig:dist_exo}), which include 55~Cnc~e whose substellar-point temperature is estimated to be about 2700~K, with zero planetary albedo. Most of the close-in small exoplanets, if they are rocky, are probably planets that have lost their primordial hydrogen-rich atmosphere due to photo-evaporation. The closer the planets are to the star, they may be bare of all volatiles, but their rocky surfaces are thought to be molten and they have secondary atmospheres vaporised from the magma due to the high temperatures. We call these rocky planets hot rocky super-Earths (hereafter HRSEs). 

Depending on their evolutionary pathways, starting as solar system type terrestrial planets or inwards migrated mini-Neptunes from beyond the ice line, they may be rocky planets, but their interior structures and compositions are mostly unknown at present. Several theoretical studies argue for the presence of not only terrestrial planets with similar interiors to those of solar system's rocky planets, but also (iron-)coreless planets \citep{Elkins-Tanton+2008}, carbon-rich planets \citep{Madhusudhan+2012,Miozzi+2018}, water-rich planets \citep{Lichtenberg+19,Zeng+19} or Ca-Al-rich planets \citep{dorn2019new}. To determine the interior composition of rocky exoplanet, atmospheric observations of HRSEs in particular allow direct constraints, because their secondary atmospheres are likely composed of materials directly vaporised from their magma ocean. 

If HRSEs are dry, they likely have atmospheres composed of rocky materials such as Na, K, Fe, Si, O, O$_2$ and SiO \citep{Schaefer+2009,Miguel+2011,Ito+2015}. On the other hand, if HRSEs have remaining volatile elements such as H, C, N, S and Cl, they likely have atmospheres composed mainly of H$_2$O and/or CO$_2$ with rocky vapours such as Na and SiO \citep{Bower+18,Schaefer+2012}.Or by HCN in case of N-dominated atmospheres \citep{Miguel2018, Zilinskas+2020} Generally, we call the former case a mineral atmosphere and the latter case a steam atmosphere. 
Thus, detection of rocky vapour would provide a definitive piece of evidence for HRSEs and their surface composition, including indications of their deeper structure. Identifying the atmospheric constituents could give constraints on the bulk composition and formation processes of the HRSEs, but this requires an atmosphere with low cloud coverage (compare report of the {\it Ariel} Chemical Working Group). Currently, no studies on the vaporised atmospheres of molten coreless, carbon-rich, or Ca-Al-rich planets exist, and we suggest that this topic should be addressed in future research. 

{\it Ariel} observations will provide clues for interior structure of HRSEs and suffice to distinguish a mineral atmosphere from a cloud-free, hydrogen-rich or water-rich atmosphere, while planets covered completely with thick clouds or with no atmosphere show flat spectra, which are similar to that of mineral atmospheres, detection of Na (0.6~$\mu$m) and K (0.8~$\mu$m) with ground-based telescopes would be helpful to distinguish a mineral atmosphere from other possibilities. Therefore, {\it Ariel} would provide information on the absence or presence of volatile elements in molten surfaces of HRSEs.

Using 55~Cancri~e as an example for the discussion, composition is a matter of speculation and needs observational constraints, but the tentative detection of  hydrogen \citep{Tsiaras+2016} indicates that some concepts may be oversimplified. 
The closeness of a planet to a star suggests that all volatile components should have been evaporated, if its interior mixing had been fast enough to rapidly supply these materials to the atmosphere \citep{Kurosaki+2014,Lopez2017}.  
Numerical studies \citep{tackley+2013,Miyagoshi+2018}, however, predict that mantle convection of a super-Earth sized terrestrial planet can be very slow.  Therefore, HRSEs' magma can retain volatile materials due to the weak interior-atmosphere interaction. The presence of abundant H  but an absence of water vapour might suggest that the atmosphere is vaporised from reduced magma retaining hydrogen and also CO and SiO \citep{Schaefer+2012}. Therefore, the detection of these gas species would lead to ascertaining the reduced magma composition and mantle convection slow enough to retain hydrogen in the interior.

\subsection{{\it Ariel} observations of terrestrial planets}
Like the other planetary types, terrestrial exoplanets are of particular interest.  
With the aim to a better understand the diversity of terrestrial planet interiors and interactions with their atmosphere, transiting planets with measured masses and radii with high accuracy are ideal targets for atmospheric follow-up characterisation. This information can then be used to infer the relation between interior, formation and evolution as indicated in Table 1. 
Although the {\it Ariel} mission is not focused on small planets, it may help to address the following questions: 
\begin{itemize}
    \item {\it Ariel} measurements will be used to determine the frequency of terrestrial planets (i.e., super-Earths) with significant atmo\-sphe\-res. 
    \item Identifying metal-rich (Mercury-like) and Earth-like planets  is important for our understanding of small planets and their potential habitability. Therefore spectra of tens of small planets could help to better characterize these planets.  
    \item The diversity of potential surface compositions of terrestrial planets is yet to be determined. {\it  Ariel} could provide insight on the surface composition of many small planets orbiting close to their stars. 
    \item {\it Ariel} will also address the issue of atmosphere escape in super-Earths. This can be used to indirectly indicate the existence of magnetic fields that prevent atmospheric loss. 
    %\item The presence of a planetary magnetic field is critical for the potential habitability of planets. Yet, it is unclear whether small exoplanets have magnetic fields. Measuring  H$_2$O and CO$_2$ in the atmospheres of terrestrial planets could indirectly indicate the existence of magnetic fields that prevent atmospheric loss, possibly  plate tectonics in old systems.
\end{itemize}

Planets that are included in the {\it Ariel} MRS list of \citet{Edwards+19} and that are of particular interest in this regard include 55Cnc~e, GJ1132~b, GJ9827~b, HD219134~b, Kepler-138~b and d, LHS1140~b and c, and the seven Trappist planets b-h. Planets without measured masses are also of interest, especially when their stellar  abundances are determined. Such planets include GJ9827~c, Kepler-444b, c, d, e, HD3167b, K2-129b. 

\section{Summary} 
Exoplanet characterisation is a key goal of exoplanet science. Measurements of the planetary mass and radius alone  are insufficient to uniquely determine the planetary composition.  
{\it Ariel}'s measurements of the atmospheric composition can break some of the degeneracy in determining the planetary compositions and improve our understanding of planets. 
%by providing information on the composition of the atmosphere. 

The statistical evaluation of the various atmosphere types can assist us to discriminate among different formation and evolutionary pathways. 
%which our current and future models predict.
The mission's results will allow us to promote/dismiss current and upcoming theoretical models using observations. 
 Linking the atmospheric composition with the bulk composition, and using the information from specific elements to further constrain the composition, evolution and formation of planets is challenging and yet to be determined.  %\cite[e.g.,][]{2019ARA&A..57..617M}}. 
Clearly, progress in theory is required in order to take full advantage of the upcoming data. 
The many targets of {\it Ariel} which include planets with various masses, host stars, and orbital properties will provide a wide view of the characteristics of exoplanets and on the connection between atmospheric and bulk composition. 

Finally, the {\it Ariel} mission is expected to significantly  improve our understanding of the interplay between planet formation, evolution and internal structure. 
%and the connection between atmospheric and bulk composition. 

\section*{Acknowledgements}
 We thank the two anonymous referees for valuable comments. 
RH acknowledges support from the  Swiss National Science Foundation
(SNSF) via grant 200020\_188460.  SCW is supported by the Research Council of Norway through its Centers of Excellence funding scheme, project number 223272 (CEED). CD acknowledges support from the Swiss National Science Foundation under grant PZ00P2\_174028. 
TL received funding from the Simons Collaboration on the Origins of Life (grant no.~611576) and the SNSF (grant no.~P2EZP2-178621). 
Parts of the work presented here was conducted within the framework of the National Centre for Competence in Research PlanetS (grant no.~51NF40-141881) supported by the Swiss National Science Foundation. This research has made use of the Exoplanet Orbit Database and the Exoplanet Data Explorer at exoplanets.org.

\end{document}